%% file: CPOP_ArXiv.tex
\documentclass[a4paper,12pt]{article}

\input{header.tex}

\usepackage{colortbl,bm}
\usepackage[ruled,vlined]{algorithm2e}
\usepackage{enumerate}
\usepackage{subfig}
\usepackage{stmaryrd}
\usepackage{natbib}

\usepackage{authblk}
\usepackage{lscape}
\usepackage{amssymb}
\usepackage[titletoc,toc,title]{appendix}

\title{Detecting changes in slope with an $L_0$ penalty}
\date{}
\author[1,2]{Robert Maidstone}
\author[1,$\dag$]{Paul Fearnhead}
\author[3]{Adam Letchford}
\affil[1]{Department of Mathematics and Statistics, Lancaster University}
\affil[2]{STOR-i Doctoral Training Centre, Lancaster University}
\affil[3]{Department of Management Science, Lancaster University}
\affil[$\dag$]{Correspondence: p.fearnhead@lancaster.ac.uk}

\begin{document}
\maketitle 

%
\begin{center}
 {\bf Abstract}
 \end{center}

 Whilst there are many approaches to detecting changes in mean for a univariate time-series, the problem of detecting multiple changes in slope has
 comparatively been ignored. Part of the reason for this is that detecting changes in slope is much more challenging. For example, simple binary segmentation
 procedures do not work for this problem, whilst efficient dynamic programming methods that work well for the change in mean problem cannot be directly used for 
 detecting changes in slope. We present a novel dynamic programming approach, CPOP, for finding the ``best'' continuous piecewise-linear fit to data. We define best based
 on a criterion that measures fit to data using the residual sum of squares, but penalises complexity based on an $L_0$ penalty on changes in slope. We show that 
 using such a criterion is more reliable at estimating changepoint locations than approaches that penalise complexity using an $L_1$ penalty. Empirically CPOP has
 good computational properties, and can analyse a time-series with over $10,000$ observations and over 100 changes in a few minutes.  Our method is used to analyse
 data on the motion of bacteria, and provides fits to the data that both have  substantially smaller residual sum of squares and are more parsimonious than two competing approaches.
 
 \section{Introduction}
\label{sec:introduction}

Changepoint detection and modelling is currently one of the most active research areas in statistics due to its importance across a wide range of applications, including: finance \cite[]{Fryzlewicz2012};
bioinformatics \cite[]{Futschik:2014,Hocking2014}; environmental science \cite[]{Killick2010}; target tracking \cite[]{Nemeth2014};  fMRI \cite[]{Aston2012}; and biochemistry \cite[]{Hotz2013} 
amongst many others. 
It appears to be increasingly important for the analysis
of large scale data streams, as it is a flexible way of modelling non-stationarity or heterogeniety in these streams. 
Changepoint detection has been identified as one of the major challenges for modern, big data applications \cite[]{Frontiers2013}.
This paper focusses on the problem of detecting changes in slope. That is, we consider data whose mean varies over time, and where we model this mean as a continuous piecewise-linear function of time.

\begin{figure}
 \begin{center}
  \includegraphics[scale=0.45]{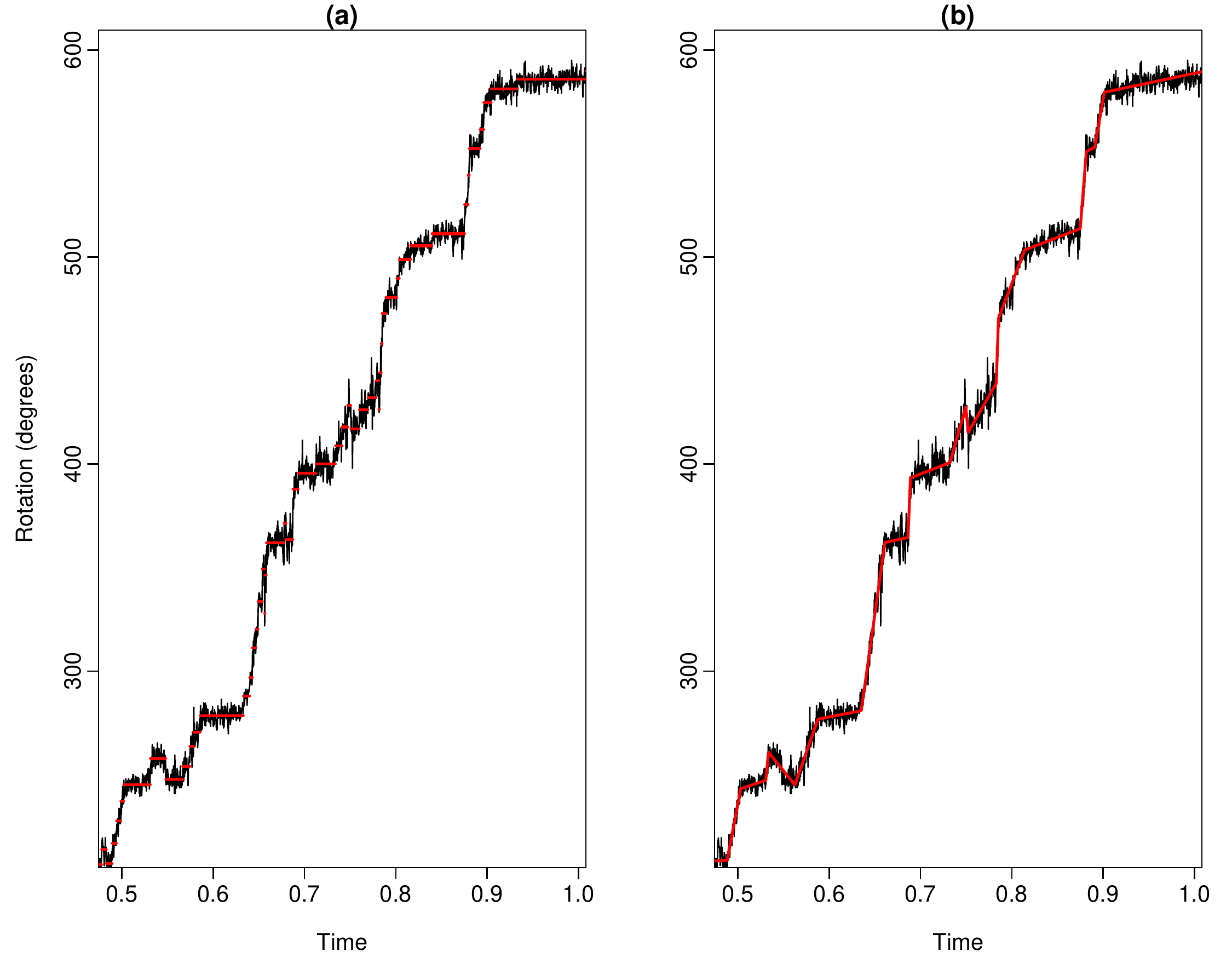}
 \end{center}
\caption{\label{Fig:Sowa1} Part of a time-series of angular position of a bacterium, taken from \cite{sowa}; best fitting piecewise constant mean (a) and continuous piecewise-linear mean (b).}
\end{figure}

To motivate this work consider the challenge of analysing data of the angular position and velocity of a bacterium, see Figure \ref{Fig:Sowa1}. 
The interest is in understanding the movement of the bacterium. The movement is driven by
the bacterial flagella, a slender thread-like structure that enables the bacteria to swim. The movement is circular, and thus the position of a bacterium
at any time point can be summarised by its angular position. The data we show comes from \cite{sowa} and was obtained
by first taking images of the bacterium at high-frequency. From these images the angular position is calculated at each time-point. The motion
is then summarised by a time-series of the amount of rotation that the bacterium has done from its initial position.

The interest from such data is in deriving understanding about the bacterial flagella motor. In particular the angular motion is characterised by stationary periods interspersed by
periods of roughly constant angular velocity. The movement tends to be, though is not exclusively, in one direction. 

\cite{sowa} analyse this data using a changepoint model, where the mean is piecewise constant. An example fit from such a model is shown in \ref{Fig:Sowa1}(a). This model is not
a natural model given the underlying physics of the application, and this can be seen in how it tries to fit periods of rotation by a number of short stationary regimes. A more natural 
model is one whereby we segment the data into periods of constant angular velocity. Such a model is equivalent to fitting a continuous piecewise-linear mean function to the data, with the slope of this function in each
segment corresponding to the angular velocity in the segment. Such a fit is shown in \ref{Fig:Sowa1}(b).

Whilst detecting changes in slope seems to be a similar statistical problem to detecting changes in mean, it is fundamentally more challenging. For example, binary segmentation approaches
\cite[]{Scott1974,Fryzlewicz2012}, which are the most popular generic approach to detecting multiple changepoints, do not work for detecting changes in slope \cite[as shown by][]{Baranowski:2016}. 
Binary segmentation iteratively applies a method for detecting a single changepoint. For change in slope problems one can show that for some underlying signals, initial estimates of changepoint locations
will tend to be midway between actual changepoint locations; binary segmentation is unable to then recover from such incorrect initial estimates.

The standard approach to detecting changes in mean is to attempt to find the ``best'' piecewise-constant mean function, where best is defined based on its fit to the data penalised by a measure of complexity of
the mean function \cite[]{Yao1988,Lavielle2000}. The most common measure of fit is through the residual sum of squares, and the most natural measure of complexity is the number of changepoints. 
The latter corresponds to an $L_0$ penalty on the change in the slope of the mean. Dynamic programming can be used to efficiently find the best segmentation of the data under such a criterion for the change in mean problem
\cite[]{Jackson2005,Killick2012a,Maidstone2014}.

Our statistical approach is to use the same framework to detect changes in slope. We aim to find the best continuous piecewise-linear mean function, where best is defined in terms of the residual sum of squares
plus a penalty that depends on the number of changepoints. However standard dynamic programming algorithms cannot be directly applied to such a problem. The assumption of continuity introduces dependencies in
the parameters associated with each segment, and these in turn violate the conditional independence structure that existing dynamic programming algorithms use. 
Detecting changes in slope under this criterion lies within a class of NP-hard problems \cite[]{weinmannandreas}. 
It is not clear to us whether our specific problem is NP-hard, but, as far as we are aware, no polynomial-time algorithm has yet been found.
Despite this, 
we present a dynamic programming algorithm that does find the best segmentation under this criterion, and has practicable computational cost -- of the order of minutes when analysing $10,000$ data points with of 
the order of 100 changepoints.

There has been earlier work on detecting changes in slope using the same or similar statistical criteria. These include \cite{Tome2004} who use an exhaustive search to find the best segmentation -- an approach that is
only feasible for very small data sets, with perhaps at most 100 to 200 data points. Alternatively, approximate solutions to the true optimal segmentation are found, for example by discretising the locations in 
time and space where changes can occur \cite[]{Goldberg2014} or by using a genetic algorithm to approximately solve the optimisation problem \cite[]{Horner1996}. As we show, our novel dynamic programming approach
is guaranteed to find the best segmentation under our criterion, and is still computationally feasible for large data sets. Empirical results suggest the expected computational cost of our algorithm is slightly worse 
than quadratic in the number of data points, and can be close to linear in situations where the number of changepoints increases linearly with the number of data points.

The outline of the paper is as follows. The next section defines the statistical criterion that we use for detecting changes in slope, and defines the optimisation problem we wish to solve in order to
find the best segmentation of the data. We present our dynamic programming algorithm, which we call CPOP, in Section \ref{sec:methodname}. We then empirically evaluate the computational and 
statistical performance of CPOP. For the latter we compare with trend-filtering \cite[]{Tibshirani2014} and the narrowest-over-threshold (NOT) approach \cite[]{Baranowski:2016}. The former involves replacing the $L_0$ penalty on changes in slope
with an $L_1$ penalty, so that we penalise mean functions based on how much, rather than the number of times, their slope changes. This makes the resulting optimisation problem convex, and hence easy to solve. However we show that whilst trend-filtering can estimate the underlying mean function well, it never performs
well at accurately detecting where the changes occur. The NOT approach is a novel version of binary segmentation that can be shown to give consistent estimation of changepoint locations
for our change in slope model. Our results show it performs well at detecting and estimating the location of the changepoints, but is less accurate than CPOP at estimating the underlying mean. 
In Section \ref{sec:bacteria} we analyse the data from Figure \ref{Fig:Sowa1}. We give statistical evidence that a change in slope model is better than fitting either a piecewise-constant or a discontinuous piecewise-linear
mean function to the data. We also show that CPOP is able to find much better fitting estimates of the mean with substantially fewer changepoints than either trend-filtering or NOT. Finally, the dynamic programming approach
we present in this paper can be applied to a larger range of changepoint problems than the change in slope problem we consider. These possible extensions are discussed in Section \ref{sec:discussion}.

\section{Model Definition}
\label{sec:normal-data}
We assume that we have data ordered by time and denote this by $\mathbf{y}=(y_1,\hdots,y_n)$. We will also use the notation that for $t\geq s$ the set of observations from time $s$ to time $t$ is $\mathbf{y}_{s:t}=(y_s,\hdots,y_t)$. If we assume that there are $m$ changepoints in the data, this will correspond to the data being split into $m+1$ distinct segments. 
We let the location of the $j$th changepoint be $\tau_j$ for $j=1,\hdots,m$, and set $\tau_0=0$ and $\tau_{m+1}=n$. The $j$th segment will consist of data points $y_{\tau_{j-1}+1},\ldots,y_{\tau_j}$. We let 
$\boldsymbol{\tau}=(\tau_0,\ldots,\tau_{m+1})$ be the set of ordered changepoints.

We consider the case of fitting a continuous piecewise linear function to the data. An example of such a fit is given in the right-hand plot of Figure~\ref{Fig:Sowa1}. 
For such a problem, changepoints will correspond to points in time where the slope of the function changes. There are a variety of ways of parameterising the linear function within each segment. Due to the
continuity constraint that we wish to enforce it is helpful to parameterise this linear function by its value at the start and its value at the end of the segment. Our continuity constraint
then requires this value for the end of one segment to be equal to the value at the start of the next segment. For the changepoint $\tau_i$ we will denote this common value as $\phi_{\tau_i}$. A continuous piecewise 
linear function is then defined by the set of changepoints, and these values of the linear function at the changes, $\phi_{\tau_i}$ for $i=0,\ldots,m+1$. As for the changepoints, we will simplify notation slightly by
letting $\boldsymbol{\phi}=(\phi_{\tau_0},\ldots,\phi_{\tau_{m+1}})$. In situations where we refer to a subset of this vector we will use the notation $\boldsymbol{\phi}_{j:k}=(\phi_{\tau_j},\ldots,\phi_{\tau_{k}})$ for 
$0\leq j \leq k \leq m+1$.
 
Under this parameterisation, we model the data as, for $i=0,\ldots,m$,
\begin{align}
\begin{array}{clc}
  & Y_t= \phi_{\tau_i}+\frac{\phi_{\tau_{i+1}}-\phi_{\tau_i}}{\tau_{i+1}-\tau_{i}}(t-\tau_i)+Z_t, &\mbox{for }t=\tau_i+1,\hdots,\tau_{i+1},\label{endpointsform}\\
\end{array}
\end{align}
where $Z_t$, for $t=1,\ldots,n$, are independent, zero-mean, random variables with common variance $\sigma^2$.

 
Our aim is to infer the set of changepoints, and the underlying piecewise linear function, from the data. Our approach to doing this is based on a penalised cost approach, using a squared-error loss function
to measure fit to the data. That is, we wish to minimise over $m$, $\boldsymbol{\tau}$, and $\boldsymbol{\phi}$,
\begin{equation} \label{eq:cost}
  \sum_{i=0}^{m} \left[\frac{1}{\sigma^2}\sum_{t=\tau_i+1}^{\tau_{i+1}} \left(y_t -  -\phi_{\tau_i}-\frac{\phi_{\tau_{i+1}}-\phi_{\tau_i}}{\tau_{i+1}-\tau_{i}}(t-\tau_i) \right)^2 + h(\tau_{i+1}-\tau_{i}) \right] + \beta m,
\end{equation}
for some suitable choice of penalty constant $\beta>0$ and segment-length penalty function $h(\cdot)$. These penalties are needed to avoid over-fitting of the data. Perhaps the most common choice of penalty is BIC 
\cite[]{Schwarz:1978}, where $\beta=2\log(n)$ and $h(s)=0$ for all segment lengths $s$. However, it has been shown that allowing the penalty to depend on segment length can improve the accuracy of 
penalised cost approaches, and
such penalties have been suggested through a modified BIC penalty \cite[]{Zhang2007} and within the minimum description length approach \cite[]{Davis2006}.
The above cost function assumes knowledge of the noise variance, $\sigma^2$.  In practice this is not known and needs to be estimated, 
for example using the Median Absolute Deviation estimator \citep{hampel1974influence}; see for example \cite{Fryzlewicz2012}.
 
We can simplify (\ref{eq:cost}) through introducing segment costs. Define the segment cost for fitting the mean of the data $\mathbf{y}_{s+1:t}$ with a linear function that starts at $\phi$ at time $s$ and
ends at $\psi$ at time $t$ as
 \begin{align}
   \mathcal{C}(\mathbf{y}_{s+1:t},\phi,\psi)=\frac{1}{\sigma^2}\sum_{j=s+1}^{t}\left(y_j-\phi-\frac{\psi-\phi}{t-s}(j-s)\right)^2.\label{costtrend}
 \end{align}
Then we wish to estimate the number and location of the changepoints, and the underlying continuous piecewise-linear function through solving the following minimisation problem:
\begin{align}
\begin{array}{lll}
   \displaystyle\min_{\boldsymbol{\tau},m,\boldsymbol{\phi}} &
   \displaystyle\left\{\sum_{i=0}^m\left[\mathcal{C}(\mathbf{y}_{\tau_i+1:\tau_{i+1}},\phi_{\tau_i},\phi_{\tau_{i+1}})+h(\tau_{i+1}-\tau_i)\right]+\beta(m+1)\right\}.&
\end{array} \label{linpro}
\end{align}

\section{Minimising the Penalised Cost}
\label{sec:methodname} 

Solving the minimisation problem in \eqref{linpro} by complete enumeration takes $\mathcal{O}(2^n)$ time and therefore is infeasible for large values of $n$. 
Below we propose a pruned dynamic programming approach to calculate the exact solution to \eqref{linpro} efficiently. This dynamic programming approach is much more complicated
than other dynamic programming algorithms used in changepoint detection as neighbouring segments share a common parameter: the end-point of the piecewise linear function for one segment is the start-point of
this function for the next segment.

Dynamic programming requires a conditional separability property. We need to be able to choose some information at time $s$ such that, conditional on this information, we can separately minimise the cost related to the 
data before and after $s$. For simpler changepoint problems, this information is just the presence of a changepoint at $s$: as conditional on this, we can separately find the best segmentation of the data before $s$
and the best segmentation of the data after $s$. For our changepoint problem, the fact that neighbouring segments share a parameter means that conditioning just on the presence of a changepoint at $s$ will no
longer give us the required separability. Instead, we will introduce a continuous-state dynamic programming algorithm which conditions on both the location of a changepoint at $s$ and the value of the function at $s$.
The idea is that given
both these pieces of information we can separately find the best segmentation of the data before $s$ and the best segmentation of the data after $s$.

\subsection{Dynamic Programming Approach} 
\label{sec:general-formulation}    

Consider segmenting the data up to time $t$, $\mathbf{y}_{1:t}$, for $t=1,\hdots,n$. When segmenting $\mathbf{y}_{1:t}$ with $k$ changepoints, $\tau_1,\hdots,\tau_k$, 
we use the notation $\tau_0=0$ and $\tau_{k+1}=t$. We define the function $f^t(\phi)$ to be the minimum penalised cost for segmenting $\mathbf{y}_{1:t}$ conditional on $\phi_t=\phi$, 
that is the fitted value at time $t$ is $\phi$. Formally this is defined as 
\begin{eqnarray}
  f^t(\phi)&=\displaystyle\min_{\boldsymbol{\tau},k,\boldsymbol{\phi}_{0:k}} 
  \displaystyle\left\{\sum_{i=0}^{k-1}\left[\mathcal{C}(\mathbf{y}_{\tau_i+1:\tau_{i+1}},\phi_{\tau_i},\phi_{\tau_{i+1}})+h(\tau_{i+1}-\tau_i)\right] \right. \nonumber \\
 &  + \left[ \mathcal{C}(\mathbf{y}_{\tau_k+1:t},\phi_{\tau_k},\phi) + h(t-\tau_k)\right]
  +\beta(k+1)\Bigg\}.  \label{eq:ft}
\end{eqnarray}
By manipulating (\ref{eq:ft}), and using the initial condition that $f^0(\phi)=0$, we can construct a dynamic programming recursion for $f^t(\phi)$.
\begin{align*}
  f^t(\phi)
&=\displaystyle\min_{\boldsymbol{\tau},k,\boldsymbol{\phi}_{0:k}} \displaystyle
\left\{\sum_{i=0}^{k-1}\left[\mathcal{C}(\mathbf{y}_{\tau_i+1:\tau_{i+1}},\phi_{\tau_i},\phi_{\tau_{i+1}})+h(\tau_{i+1}-\tau_i)\right]+\beta k\right.\\
&+\mathcal{C}(\mathbf{y}_{\tau_k+1:t},\phi_{\tau_k},\phi_t)+h(t-\tau_k)+ \beta\Bigg\},\\
&=\min_{\phi',s}\left\{\displaystyle\min_{\boldsymbol{\tau}_{0:k-1},k,\boldsymbol{\phi}_{0:k-1}} \displaystyle
\left\{\sum_{i=0}^{k-2}\left[\mathcal{C}(\mathbf{y}_{\tau_i+1:\tau_{i+1}},\phi_{\tau_i},\phi_{\tau_{i+1}})+h(\tau_{i+1}-\tau_i)\right]+ \right.\right.\\
&+\mathcal{C}(\mathbf{y}_{\tau_{k-1}+1:s},\phi_{\tau_{k-1}},\phi')+h(s-\tau_{k-1})+\beta k\Bigg\}+\mathcal{C}(\mathbf{y}_{s+1:t},\phi',\phi)+h(t-s)+\beta\Bigg\},\\
&=\min_{\phi',s}\left\{f^{s}(\phi')+\mathcal{C}(y_{s+1:t},\phi',\phi)+h(t-s)+\beta\right\}.
\end{align*}
The idea is that we split the minimisation into first minimising over the time of the most recent changepoint and the fitted value at that changepoint, and then minimising over the earlier changepoints and fitted values. 
On the third line we let $s$ denote the time of the most recent changepoint, and $\phi'$ the fitted value at $s$. The inner minimisation is over the number of changepoints, 
the locations of those changepoints prior to $s$, and the fitted values at the changepoints prior to $s$. This inner minimisation gives the minimum penalised cost for 
segmenting $\mathbf{y}_{1:s}$ conditional on $\phi_s = \phi'$, which is $f^s(\phi')$. This recursion is similar to that derived for Optimal Partitioning. However for Optimal Partitioning we just needed to store 
a scalar value for each $t = 1,\hdots n$. Here we need to store functions of the continuous parameter $\phi$ for each value of $t$.



To store $f^t(\phi)$ we will write it as the point-wise minimum of a set of cost functions of $\phi$, each of which corresponds to a different vector of changepoints, $\boldsymbol{\tau}$. 
We define each of these functions $f_{\boldsymbol{\tau}}^t(\phi)$ as the minimum cost of segmenting $\mathbf{y}_{1:t}$ with changepoints at $\boldsymbol{\tau}=\tau_1,\hdots,\tau_k$ and 
fitted value at time $t$ being $\phi$:
\begin{eqnarray}
f_{\boldsymbol{\tau}}^t(\phi)&=    \displaystyle\min_{\boldsymbol{\phi}_{0:k}}
\displaystyle\left\{\sum_{i=0}^{k-1}\left[\mathcal{C}(\mathbf{y}_{\tau_i+1:\tau_{i+1}},\phi_{\tau_i},\phi_{\tau_{i+1}})+h(\tau_{i+1}-\tau_i)\right]\right. \nonumber \\
&+\mathcal{C}(\mathbf{y}_{\tau_k+1:t},\phi_{\tau_k},\phi)+h(t-\tau_k) +\beta(k+1)\Bigg\}. \label{eq:1}
\end{eqnarray}
Then $f^t(\phi)$ is the point-wise minimum of these functions,
\begin{align}
  \label{eq:4}
  f^t(\phi)=\min_{\boldsymbol{\tau}\in \mathcal{T}_t}f_{\boldsymbol{\tau}}^t(\phi),
\end{align}
where we define $\mathcal{T}_t$ to be the set of all possible changepoint vectors at time $t$. 


Each of the above functions, $f_{\boldsymbol{\tau}}^t(\phi)$, is a quadratic in $\phi$ and thus can be represented by a vector of length 3, 
with the terms in this vector denoting the co-efficients of the quadratic. We can calculate the co-efficients recursively using
\begin{align}
  \label{eq:3}
  f_{\boldsymbol{\tau}}^t(\phi)=\min_{\phi'}\left\{f_{\tau_1,\hdots,\tau_{k-1}}^{\tau_k}(\phi')+\mathcal{C}(y_{\tau_k+1:t},\phi',\phi)+h(t-\tau_{k})+\beta\right\}.
\end{align}
Further details are given in   Appendix \ref{sec:coupdate}. Therefore we can iteratively compute these functions and thus calculate $f^n(\phi)$. 

We then calculate the optimal segmentation of $\mathbf{y}_{1:n}$ by minimising $f^n(\phi)$ over $\phi$. The value of $\boldsymbol{\tau}$ that achieves the minimum value will be the optimal segmentation. 
This approach, however, is computationally expensive; both in time, $\mathcal{O}(n2^n)$, and space needed to store the functions, $\mathcal{O}(2^n)$. 
To obtain a practicable algorithm we have to use pruning ideas to reduce the number of changepoint vectors, and corresponding functions $f_{\boldsymbol{\tau}}^t(\phi)$, that we need to store.
There are two ways in which this can be achieved: functional pruning and inequality based pruning. 
In both cases they are able to remove changepoint vectors whilst still maintaining the guarantee that the resulting algorithm will find the true minimum of the optimisation problem (\ref{eq:cost}).

\subsection{Functional Pruning}
\label{sec:functional-pruning}


One way we can prune these candidate changepoint vectors from the minimisation problem is when they can be shown to be dominated by other vectors for any given value of $\phi$. Similar approaches are found in \cite{Rigaill2010} and \cite{Maidstone2014} for independent segment models and is known as \emph{functional pruning}.

In Theorem~\ref{thr:fp} we show how if a candidate changepoint vector, $\boldsymbol{\tau}$ is not optimal at time $s$ for any value of $\phi$, then the related candidate changepoint vector $(\boldsymbol{\tau},s)$ (the concatenation of $\boldsymbol{\tau}$ and $s$) is not optimal for any value of $\phi$ at time $t$ where $t>s$. If this is the case, the vector $(\boldsymbol{\tau},s)$ can be pruned from the candidate changepoint set.

First we define the set $\overset{*}{\mathcal{T}}_t$ as the set of changepoint vectors that are optimal for some $\phi$ at time $t$
\begin{align}
  \overset{*}{\mathcal{T}}_t=\left\{\boldsymbol{\tau}\in\mathcal{T}_t:f^t(\phi)=f_{\boldsymbol{\tau}}^t(\phi), \mbox{ for some }\phi\in(-\infty,\infty)\right\},\label{eq:tstardef}
\end{align}
where $\mathcal{T}_t$ is the set of all possible changepoint vectors at time $t$. If a candidate vector $\boldsymbol{\tau}$ is not in this set at time $s$ then the related candidate vector 
$(\boldsymbol{\tau},s)$ is not in the set at time $t$. This means that at time $t$ we will need to store only the functions $f_{\boldsymbol{\tau}}^t(\phi)$ 
corresponding to segmentations that are in $\overset{*}{\mathcal{T}}_t$. 

\begin{theorem}
  If $\boldsymbol{\tau}\notin\overset{*}{\mathcal{T}}_s$ then $(\boldsymbol{\tau},s)\notin\overset{*}{\mathcal{T}}_t$ for all $t>s$.\label{thr:fp}
\end{theorem}
{\bf Proof:} See Appendix \ref{App:Proof}.

The key to an efficient algorithm will be a way of efficiently calculating $\overset{*}{\mathcal{T}}_t$. We can use the above theorem to help us do this. From Theorem \ref{thr:fp} we can define a set
\begin{align}
\hat{\mathcal{T}}_t= \left\{(\boldsymbol{\tau},s): s\in\{0,\hdots,t-1\}, \boldsymbol{\tau}\in\overset{*}{\mathcal{T}}_s\right\},\label{supsets}
\end{align}
and we will have that $\hat{\mathcal{T}}_t\supseteq\overset{*}{\mathcal{T}}_t$. So assume that we have calculated the sets $\overset{*}{\mathcal{T}}_s$ for $s=0,\hdots,t-1$. We can calculate $f_{\boldsymbol{\tau}}^t(\phi)$ only for $\boldsymbol{\tau}\in\hat{\mathcal{T}}$. When calculating $f^t(\phi)$, as defined by (\ref{eq:4}), we can just minimise over the set of changepoint vectors in $\hat{\mathcal{T}}_t$ rather than the full set. Furthermore we can calculate which of the sets of changepoints in $\hat{\mathcal{T}}_t$ contribute to this minimum and remove those that do not contribute. The remaining sets of changepoints define $\overset{*}{\mathcal{T}}_t$.

To find out which sets of changepoints, $\boldsymbol{\tau}$, contribute to the minimisation of (\ref{eq:4}) we store the interval (or set of intervals) of $\phi$ space for which it is optimal. We define this interval as follows
\begin{align}
  Int^t_{\boldsymbol{\tau}}=\left\{\phi:f_{\boldsymbol{\tau}}^t(\phi)=\displaystyle\min_{\boldsymbol{\tau}'\in\hat{\mathcal{T}}_t}f_{\boldsymbol{\tau}'}^t(\phi)\right\}.\label{eq:setcalc}
\end{align}

 For a given $t$ the union of these intervals over $\boldsymbol{\tau}$ is just the real line (as for a given $\phi$ at least one changepoint vector $\boldsymbol{\tau}$ corresponds to the optimal segmentation). 
 Using this we can derive a simple algorithm for updating these intervals which involves a search over the real line, recursively finding the function, and associated interval, which is optimal as we increase $\phi$ from $-\infty$ to $\infty$.   This method is given in full in Algorithm~\ref{algo_calcsettau}, and there is a detailed explanation in Appendix \ref{App:Alg}.

Having calculated $Int^t_{\boldsymbol{\tau}}$ for all $\boldsymbol{\tau}\in\hat{\mathcal{T}}$ we can use these to calculate $\overset{*}{\mathcal{T}}$. We remove $\boldsymbol{\tau}$ from $\hat{\mathcal{T}}$ if $Int^t_{\boldsymbol{\tau}}=\emptyset$ and after doing this for all $\boldsymbol{\tau}\in\hat{\mathcal{T}}$ we are left with precisely those values of $\boldsymbol{\tau}$ which make up $\overset{*}{\mathcal{T}}$. This is used to recursively calculate $\hat{\mathcal{T}}_{t+1}$
\begin{align}
 \hat{\mathcal{T}}_{t+1}=\hat{\mathcal{T}}_t\cup\left\{(\boldsymbol{\tau},t):\boldsymbol{\tau}\in\overset{*}{\mathcal{T}}_t\right\}.\label{eq:Ttplus1}
\end{align}



\subsection{Inequality Based Pruning}
\label{sec:pelt-like-pruning}

A further way pruning can be used to speed up the dynamic programming algorithm is by applying \emph{inequality based pruning} \cite[]{Maidstone2014}, a similar idea to the pruning
step used in the PELT algorithm \cite[]{Killick2012a}. This pruning is based on the following result.
\begin{theorem}
  \label{thr:3}
  Define $K=2\beta+h(1)+h(n)$.
If $h(\cdot)$ is a non-negative, non-decreasing function and
if for some $\boldsymbol{\tau}$,
\begin{align}
  \min_{\phi}f_{\boldsymbol{\tau}}^t(\phi) >  \min_{\phi'}\left[f^{t}(\phi')\right]+K,\label{eq:hold}
\end{align}
then at any future time $T$, the set of changepoints $\boldsymbol{\tau}$ can never be optimal for the data $\mathbf{y}_{1:T}$.
\end{theorem}
{\bf Proof:} See Appendix \ref{App:Proof}.

This result states that for any candidate changepoint vector,  if the best cost at time $t$ is worse than the best cost over all changepoint vectors plus $K$,
we can show that the candidate is sub-optimal at all future times as well.
In Section~\ref{sec:functional-pruning} we considered candidate changepoints vectors that belonged to the set $\hat{\mathcal{T}}_t$, and updated the related cost functions. 
We then used functional pruning to reduce this set to only those values that are optimal for some value of $\phi$, namely the set $\overset{*}{\mathcal{T}}$. 
Using Theorem \ref{thr:3} we can reduce the size of $\hat{\mathcal{T}}_{t+1}$ before the cost functions are updated, discarding candidates from the set if (\ref{eq:hold}) is true. 
As this reduces the size of the set $\hat{\mathcal{T}}_t$, it also reduces the computational cost of the algorithm.


Both pruning steps can be used to restrict the set of candidate changepoint vectors that the dynamic program is run over. We call the resulting algorithm CPOP, for Continuous-piecewise-linear Pruned Optimal Partitioning. 
The pseudocode for the full method with these pruning steps is outlined in Algorithm~\ref{algo_optpiecewise} in the Appendix.

\subsection{Computational Cost of CPOP} \label{sec:CompCost}

The computational cost of the CPOP algorithm depends crucially on the size of $\overset{*}{\mathcal{T}}_t$ and $\hat{\mathcal{T}}_{t}$. Denote the size of each set by $|\overset{*}{\mathcal{T}}_t|$ and $| \hat{\mathcal{T}}_{t}|$ respectively. For iteration $t$ of the CPOP, 
the cost of calculating the quadratics, $f^t_{\boldmath{\tau}}(\phi)$, associated with each $\boldsymbol{\tau}\in\hat{\mathcal{T}}_{t}$, will be linear in $|\hat{\mathcal{T}}_{t}|$. The cost
of calculating $Int_{\boldsymbol{\tau}}^t$, the intervals of $\phi$ for which each quadratic is optimal, will have a cost that is of the order of $|\hat{\mathcal{T}}_{t}|$ times the number of disjoint intervals
that contribute to the set of $Int_{\boldsymbol{\tau}}^t$. We believe the number of such intervals increases linearly in $|\overset{*}{\mathcal{T}}_t|$. 
 Note that without the inequality-based pruning we have $|\hat{\mathcal{T}}_{t}|=\sum_{s=1}^{t-1} |\overset{*}{\mathcal{T}}_s|$.

\begin{figure}
  \centering  
 \makebox[\textwidth][c]{
\subfloat[]{
\includegraphics[trim=2 1 2 1,width=7cm]{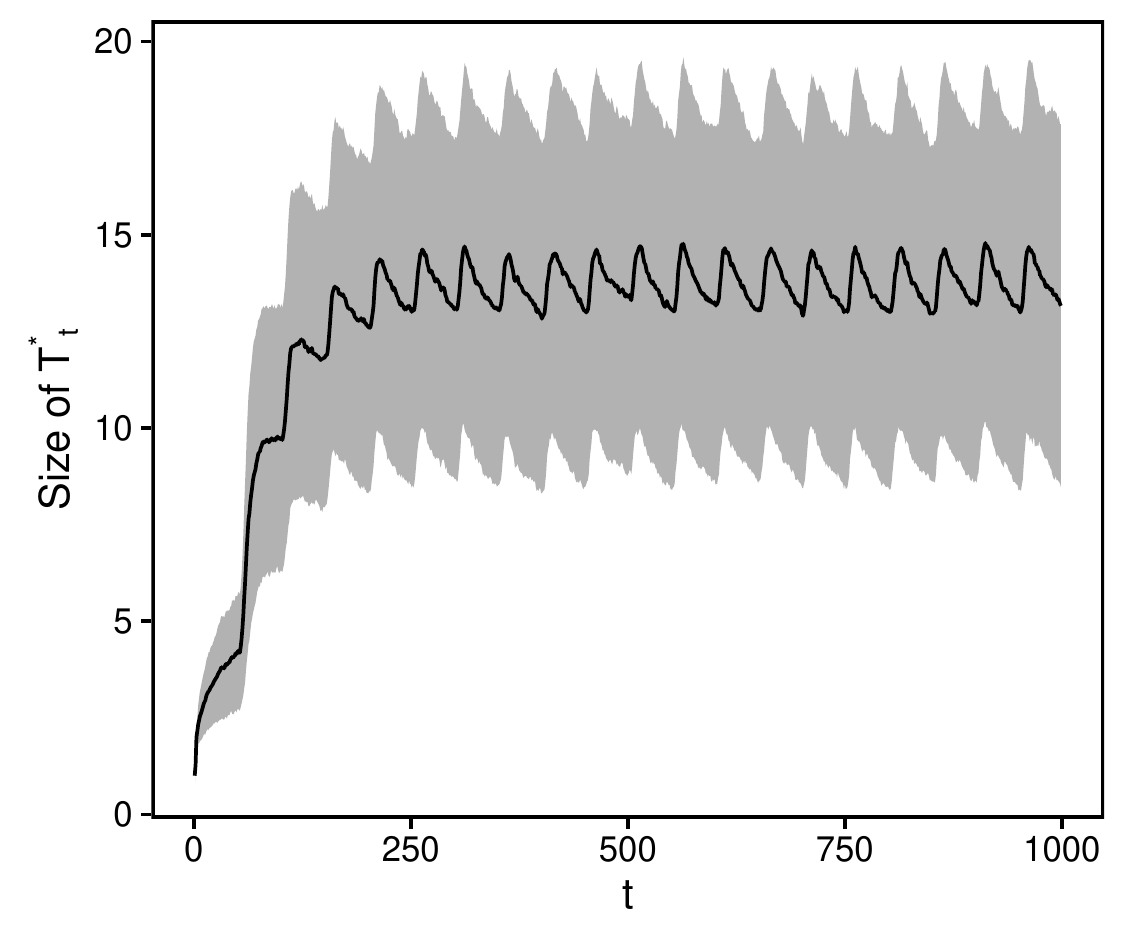}
}\hspace{5pt} 
 \subfloat[]{
\includegraphics[trim=2 1 2 1,width=7cm]{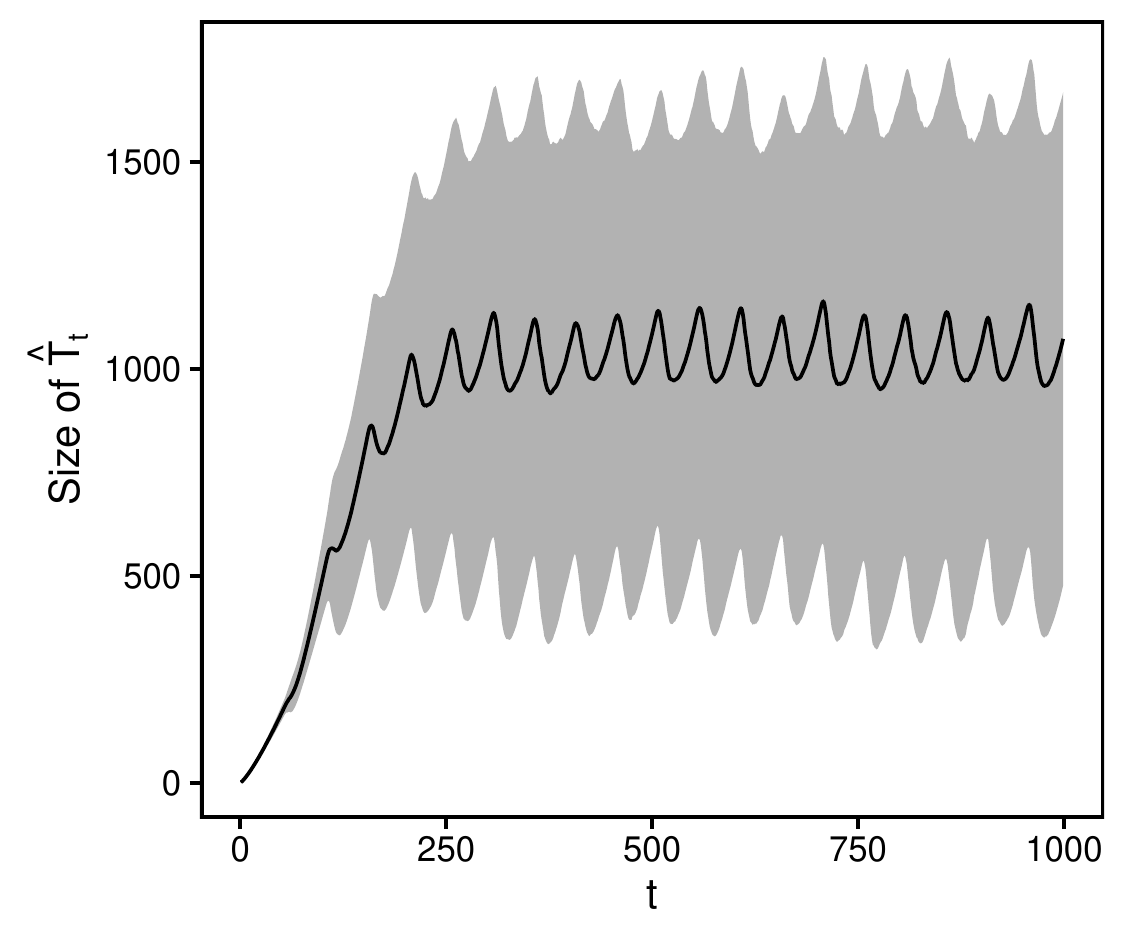} 
}}
\makebox[\textwidth][c]{
\subfloat[]{
\includegraphics[trim=2 1 2 1,width=7cm]{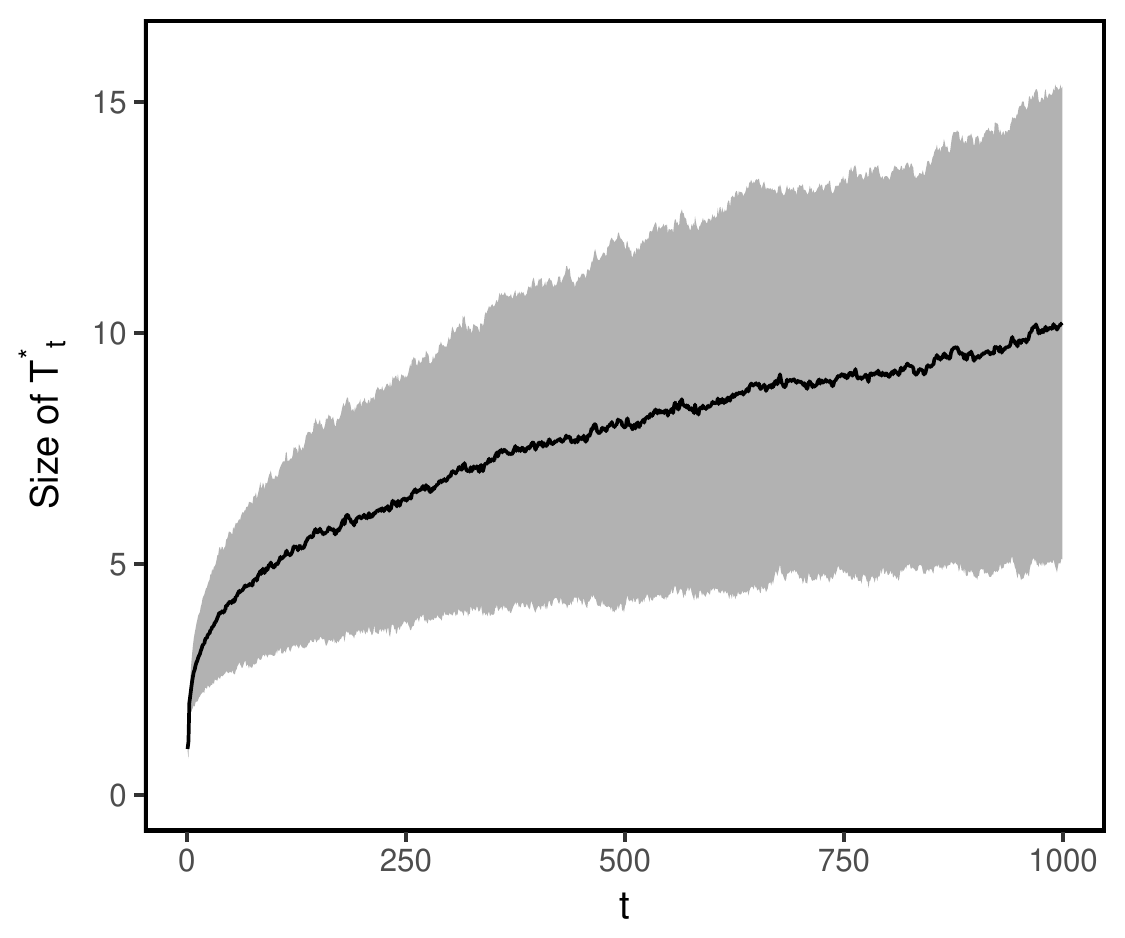}
}\hspace{5pt} 
 \subfloat[]{
\includegraphics[trim=2 1 2 1,width=7cm]{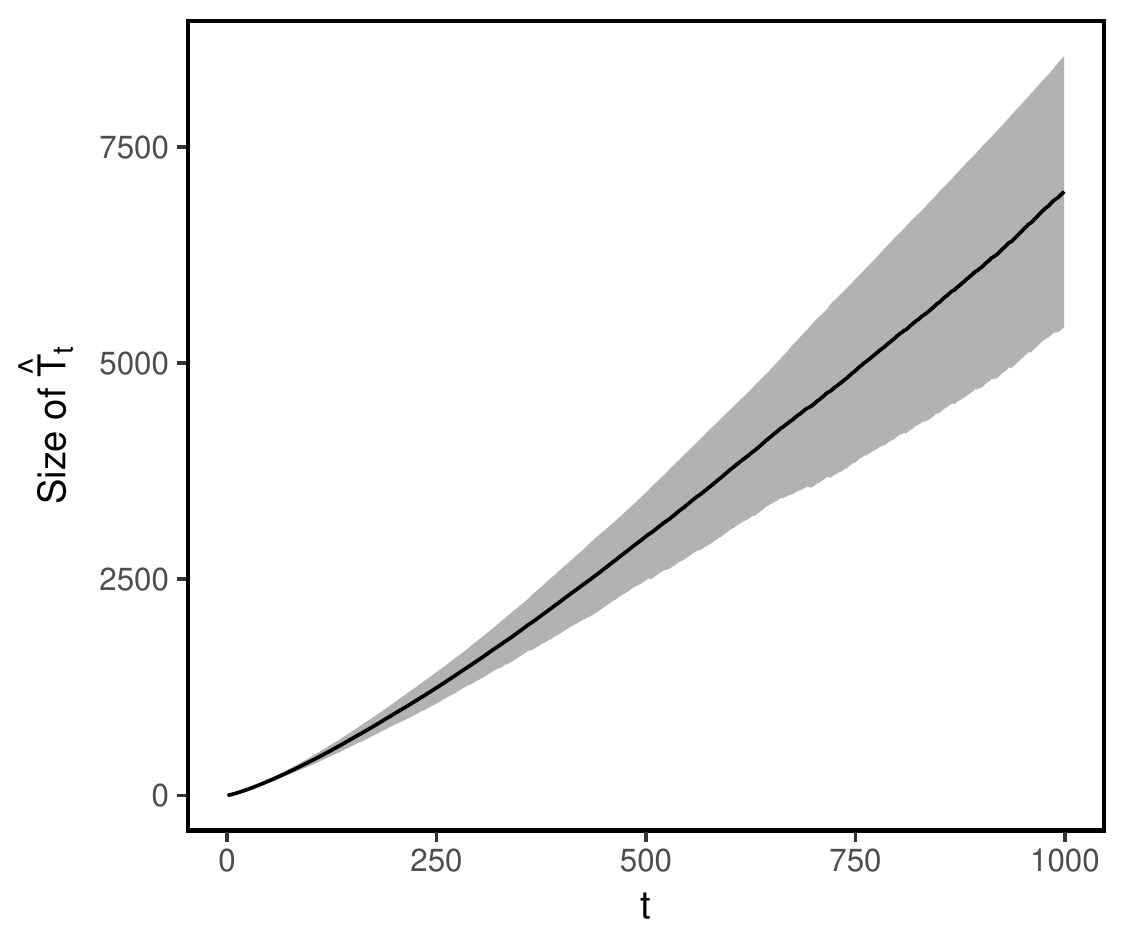}
}
}
 \caption{\label{Fig:CompCost} Size of $\overset{*}{\mathcal{T}}_t$ (left-hand column) and $ \hat{\mathcal{T}}_t$ (right-hand column) as a function of $t$ for data simulated with $m=19$ changepoints (top row) and 
 no changepoints (bottom row). Lines show the average size, and shaded regions show plus or minus 1 standard deviation. Results are based on the analysis of 1000 data sets in each case.}
\end{figure}

To investigate empirically how the size of these sets increase with $t$, and what the resulting computational cost of CPOP is, we analysed simulated data sets of different sizes, $n$, 
and with different numbers of changepoints, $m$. For a given choice of $n$ and $m$ we set the changepoints to be equally spaced, and simulated the value of the underlying mean function at the each changepoint as
an independent draw from a Gaussian distribution with variance 4. We then simulated data by adding independent, identically distributed standard Gaussian noise to this mean function at each time-point. We present
results for $n=1000$ and for both many changepoints, $m=19$, and no changepoint, $m=0$ in Figure \ref{Fig:CompCost} (qualitatively similar results were obtained for other values of $n$ and $m$). 

Without pruning, the value of $ |\overset{*}{\mathcal{T}}_t|$ would increase exponentially with $t$. However we see that in both cases $ |\overset{*}{\mathcal{T}}_t|$ remains small for all $t$, with the
average values always less than 20. 

The behaviour of $|\overset{*}{\mathcal{T}}_t|$ is different for the two cases. For the no changepoint case, the size of this set increases linearly with $t$. For the many changepoint case the size initially increases
linearly but then appears roughly constant. The reason for this is that the inequality based pruning of Section \ref{sec:pelt-like-pruning} is able to prune many of the segmentations in $|\overset{*}{\mathcal{T}}_t|$
that have a most recent changepoint which is a long-time prior to the actual most recent changepoint \cite[see][for a similar effect of this type of pruning]{Killick2012a}. This reduces the size of
$|\overset{*}{\mathcal{T}}_t|$ substantially when there are many changepoints, whereas the inequality based pruning has almost no effect for the case where there are no changepoints. 


\begin{figure}[t]
   \centering  
 \makebox[\textwidth][c]{
\subfloat[]{
  \includegraphics[trim=2 1 2 1,width=7cm]{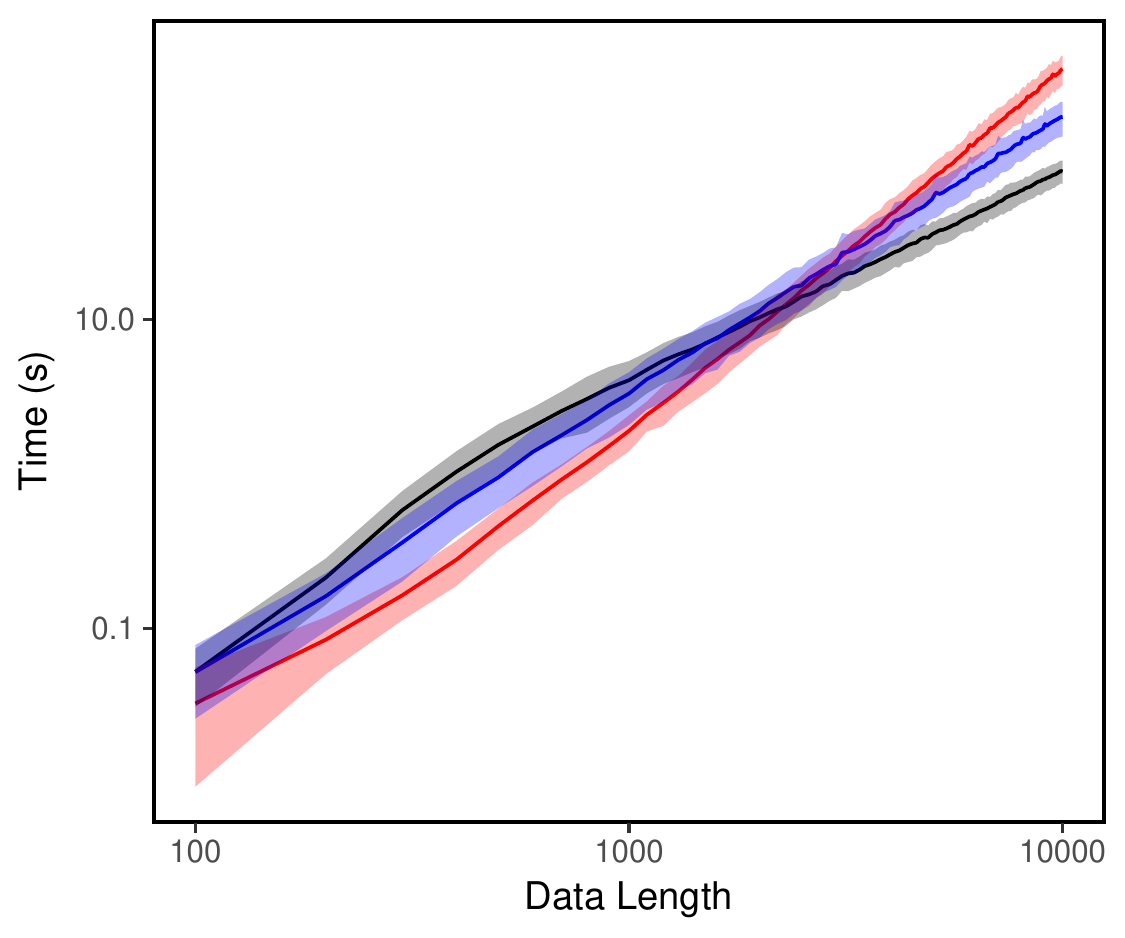}
}\hspace{5pt} 
 \subfloat[]{
 
\includegraphics[trim=2 1 2 1,width=7cm]{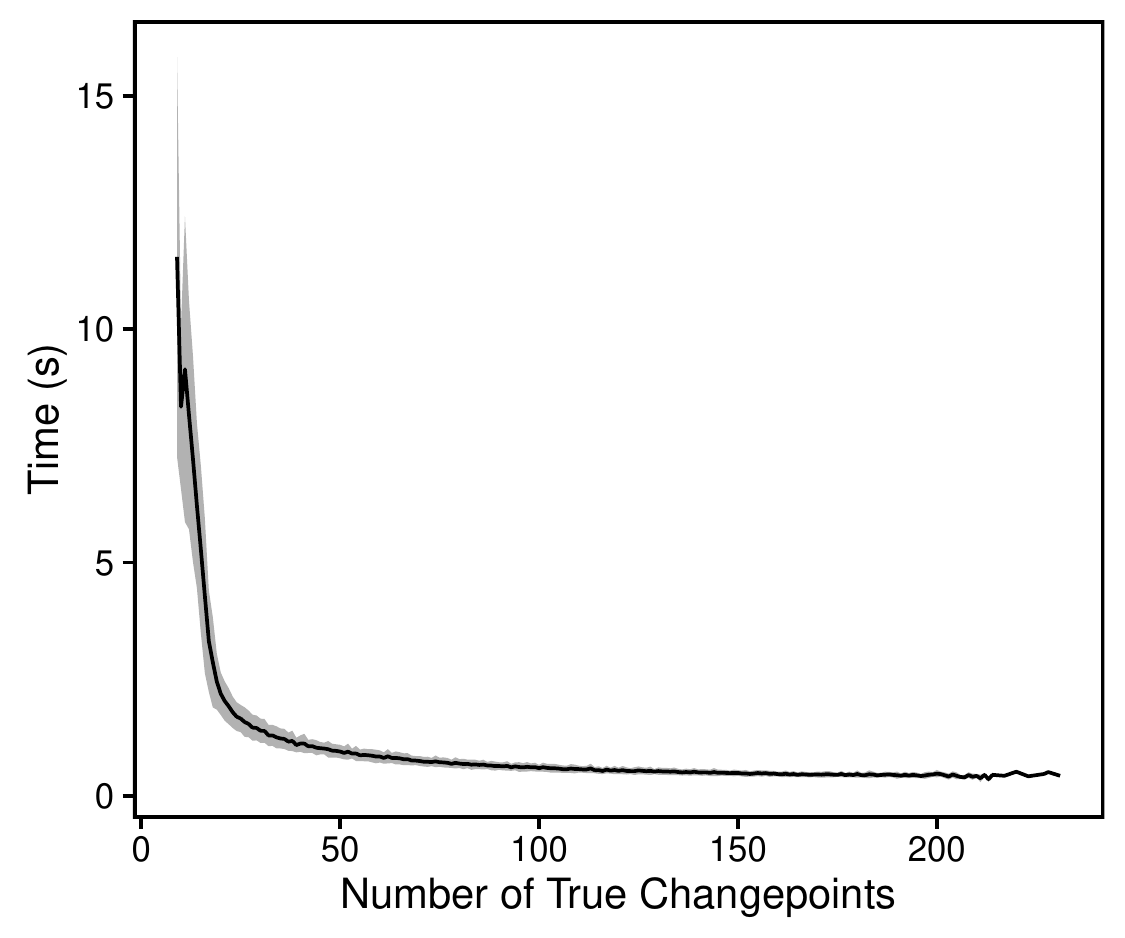}
}} 
 \caption{\label{Fig:CompCost2} Computational cost, in seconds, of CPOP as we increase $n$ (left-hand plot) and for fixed numbers of observations, $n=1000$, but increasing numbers of changepoints (right-hand plot). For the former case we have
 used a log-scale on both axes, and we give  average computational cost for three regimes for the number of changepoints, $m$: a fixed number of changepoints, $m=50$ (red); a linearly increasing number of changepoints, $m=n/50$ (black); and $m=\lfloor\sqrt{n}\rfloor$ (blue).
 Lines show the average computational cost, and shaded regions show plus or minus 1 standard deviation. 
 }
 \end{figure}

We also empirically investigated the overall computational cost of CPOP for different sizes of data set, $n$, and different numbers of changepoints, $m$. 
Figure \ref{Fig:CompCost2} shows the average time for CPOP. The first plot is of computational cost against $n$ for three different regimes for $m$. 
For each of the three regimes we see a roughly linear relationship between the log computational cost and $\log(n)$. The slopes of these lines vary between 1.3 for the fixed $m$ regime and 2.3 for the regime where $m$ 
increases linearly with $n$. These suggest computational cost grows like $n^{1.3}$ and $n^{2.3}$ respectively. This is consistent with the second plot, which shows that for fixed $n$ the computational cost 
decreases with increasing $m$.

\section{Statistical Performance of CPOP} \label{sec:Results}

We now look empirically at the statistical performance of CPOP, and compare this with two other 
methods for fitting a continuous piecewise-linear mean function to data and detecting the locations where the slope of this function changes. 

The most common, general, approach for detecting changes is to use binary segmentation \cite[]{Scott1974}, but as mentioned in the introduction  binary segmentation does not work for this problem: there are examples where even if you observed 
the underlying mean function without noise, binary segmentation would not correctly identify the changepoints. 

To overcome this, \cite{Baranowski:2016}, present the {\it narrowest-over-threshold} algorithm, henceforth called the NOT algorithm. This algorithm proceeds by (i) taking a pre-specified number, $M$, 
of intervals of data, $\mathbf{y}_{s_i:t_i}$ say; (ii) performing a generalised likelihood ratio test for a change in slope on each
$\mathbf{y}_{s_i:t_i}$; (iii) keeping all intervals for which the test-statistic is above some pre-specified threshold; (iv) ordering these intervals, with the shortest interval first and the longest last; (v) running down this list in order, adding changepoints at each of the inferred changepoint locations for an interval providing that interval does not contain any previously inferred changepoints. The idea of the algorithm is that by concentrating on the smallest intervals in (iv), these will be intervals that are likely to have at most one actual changepoint, and hence the inferred changepoint in step (v) should be close in position to this actual changepoint. 

In practice, NOT is run for a continuous range of thresholds in step (iii). This will produce a set of different segmentations of the data. The segmentation that is then chosen is the one that minimises 
the BIC for a model where the residuals are independent Gaussian with unknown variance $\sigma^2$. 
For a segmentation with $m$ changepoints at locations $\boldsymbol{\tau}$, the BIC corresponds to the minimum, over $\boldsymbol{\phi}$ of  
\[
n \log\left( \frac{1}{n} \sum_{i=0}^{m} \left[\sum_{t=\tau_i+1}^{\tau_{i+1}} \left(y_t -  \frac{\phi_{\tau_{i+1}}-\phi_{\tau_i}}{\tau_{i+1}-\tau_{i}}(t-\tau_i) \right)^2 \right]\right) + 2m\log n.
\]
This is closely related to our criterion (\ref{eq:cost}) with the BIC penalty, except for the assumption of unknown variance, and the fact that this criterion is only minimised over the 
set of segmentations found by the NOT algorithm. One advantage of this approach is that it avoids the need to have an estimate of 
$\sigma$.

The other approach we compare to is  the trend-filtering algorithm \citep{Kim2009}. Trend-filtering aims to minimise the residual sum of squares of the fitted continuous piecewise-linear mean, but with an $L_1$ penalty on how the slope changes. Again, this is closely related to our criterion (\ref{eq:cost}), except we use an $L_0$ penalty on the changes in slope.

Trend-filtering requires a choice of penalty, in the same way that we need to choose the penalty $\beta$ in (\ref{eq:cost}). To mimic the approach of NOT we use a BIC type approach. This involves running
the trend-filtering algorithm for a discrete set of penalty values. For a given penalty value, trend-filtering will output an estimate of the mean at each time point. From this we can infer the changepoint locations as the points where the estimated mean has a change in slope. We evaluate the output from each run of the trend-filtering algorithm using BIC.
If the estimated mean is $\hat{\phi}_{1:n}$, and this has $m$ changes in slope, then using the fact that for trend-filtering a segmentation with $m$ changes in slope has an 
effective degrees of freedom that is $m+2$ \citep{Tibshirani2014},
the BIC value is
\[
\frac{1}{\sigma^2}\left( \sum_{t=1}^{n} [y_t-\hat{\phi}_t]^2\right) + (m+2)\log(n).
\]
Other approaches, including fitting a change in mean to differenced data and ignoring the continuity constraint when detecting changepoints, are considered in \cite{Maidstone:thesis}. 
However these all perform  much worse, across all measures of accuracy, than the three approaches we compare here.

In the comparisons below we implement CPOP for minimising (\ref{eq:cost}) with the BIC penalty. We use the {\texttt{not}} R-package to implement NOT, 
and the code available from \url{http://stanford.edu/~boyd/l1_tf} to implement trend-filtering. For NOT we set the number of intervals, $M$ in step (i) of the algorithm above, to $10^5$. This is larger than recommended in \cite{Baranowski:2016}, but we
found it gave slightly better results than the default choice of $10^4$ intervals.
For trend-filtering and CPOP we need an estimate of the variance of the residuals.  Within a segment, the variance of the second differences of the data is easily shown to be 6 times the 
variance of the residuals. Thus we take second differences of the
data, and take one-sixth of the median-absolute-deviation estimator of the variance of these second differences. Of course, being heuristic methods, both NOT and trend-filtering are much faster algorithms than CPOP. 
Across all the scenarios we considered, trend-filtering and NOT ran in a few seconds, whereas CPOP took between tens of seconds to a few minutes.

\begin{figure}
 \begin{center}
  \includegraphics[scale=0.6]{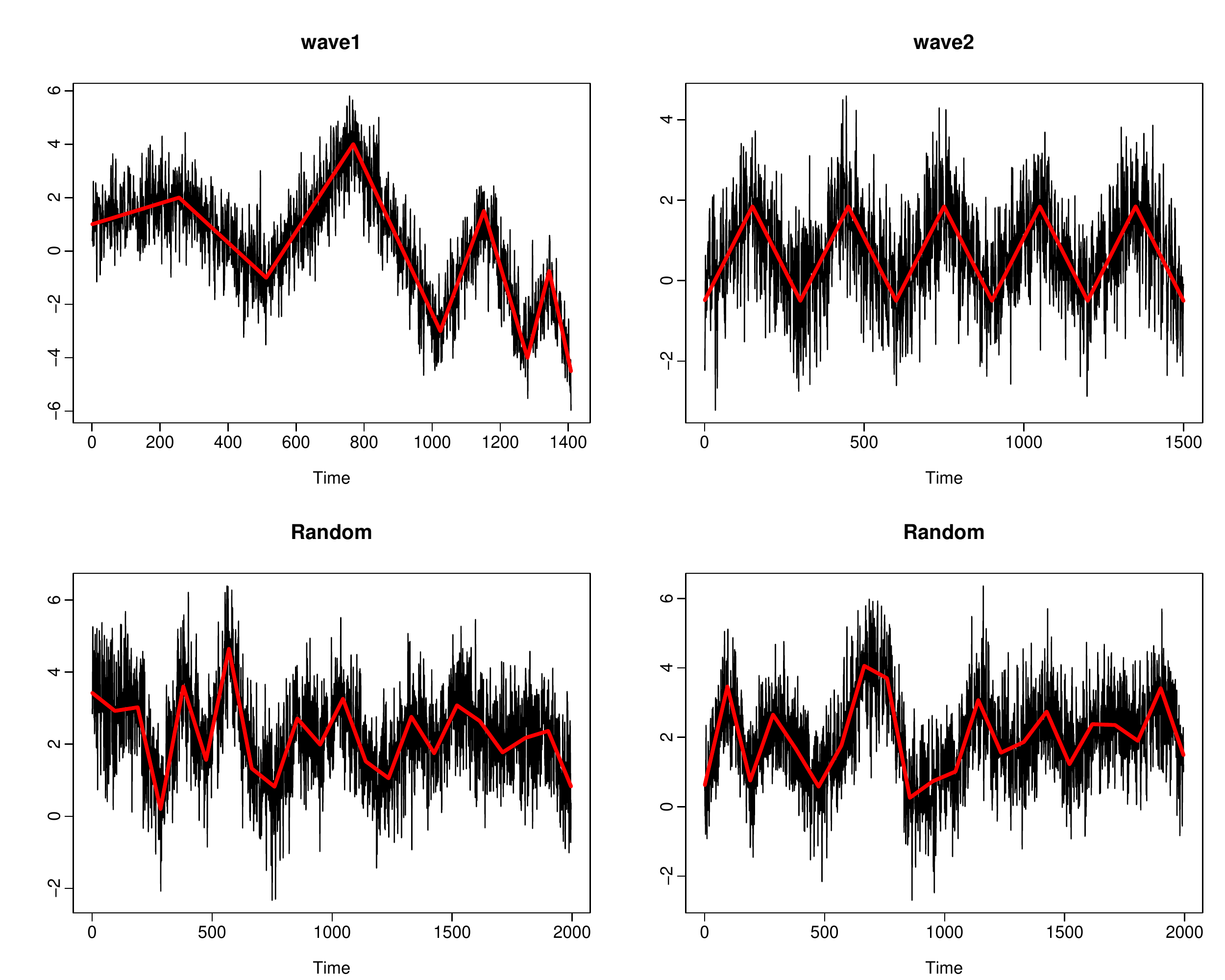}
 \end{center}
 \caption{\label{Fig:Scenarios}
 Example data from the three simulation scenarios we considered: {\texttt{wave1}} and {\texttt{wave2}} (top row) are taken from \cite{Baranowski:2016} and the shape of the mean function is fixed in these scenarios. For the {\texttt{Random}} scenario (bottom
 row), the form of the mean is random, and we give two example realisations.
 }
\end{figure}

The three scenarios that we compared the methods on are shown in Figure \ref{Fig:Scenarios}. The first two of these, {\texttt{wave1}} and {\texttt{wave2}}, are taken from \cite{Baranowski:2016}. These two
scenarios have a fixed mean function. We consider extensions of these two scenarios with higher-frequency observations for {\texttt{wave1}}, where we have twice or four times as many observations within
each segment; and longer time-series for {\texttt{wave2}}, where we have 20 or 40 segments, each of 150 observations, rather than just 10. In the third scenario, which we call {\texttt{Random}}, we simulate the underlying
mean for each data set. This setting has segments of equal length, but the value of the mean function at the start/end of each segment is simulated from a Gaussian distribution with variance 4. 
For this setting we will consider varying both the number of data points and the number of changepoints. In all cases that data is obtained by adding independent standard Gaussian noise to the mean.

Following \cite{Baranowski:2016}, for {\texttt{wave1}} and {\texttt{wave2}} we compare methods using the mean square error (MSE) of the estimates of the mean, and using a scaled Hausdorff distance, $d_H$, to measure accuracy of the changepoint locations.
This distance is defined as
\[
 d_H=\frac{1}{n_s} \max\left\{ \max_j \min_k |\tau_j-\hat{\tau}_k|, \max_k \min_j |\tau_j-\hat{\tau}_k|  \right\},
\]
where $\hat{\tau}_k$ are the estimated changepoint locations, $\tau_j$ the true changepoint locations, and $n_s$ the length of the largest segment. The idea is that for each true change we find the closest estimated
changepoint, and for each estimated changepoint we find the closest true changepoint. We then calculate the distance between each of these pairs of changepoints, and $d_H$ is set to the largest of these distances divided 
by the length of the longest segment. The smaller $d_H$ the better the estimates of the changespoints, with $d_H=0$ meaning that all changepoints are detected without error, and no other changepoints are estimated.

\begin{figure}
 \begin{center}
  \includegraphics[scale=0.6]{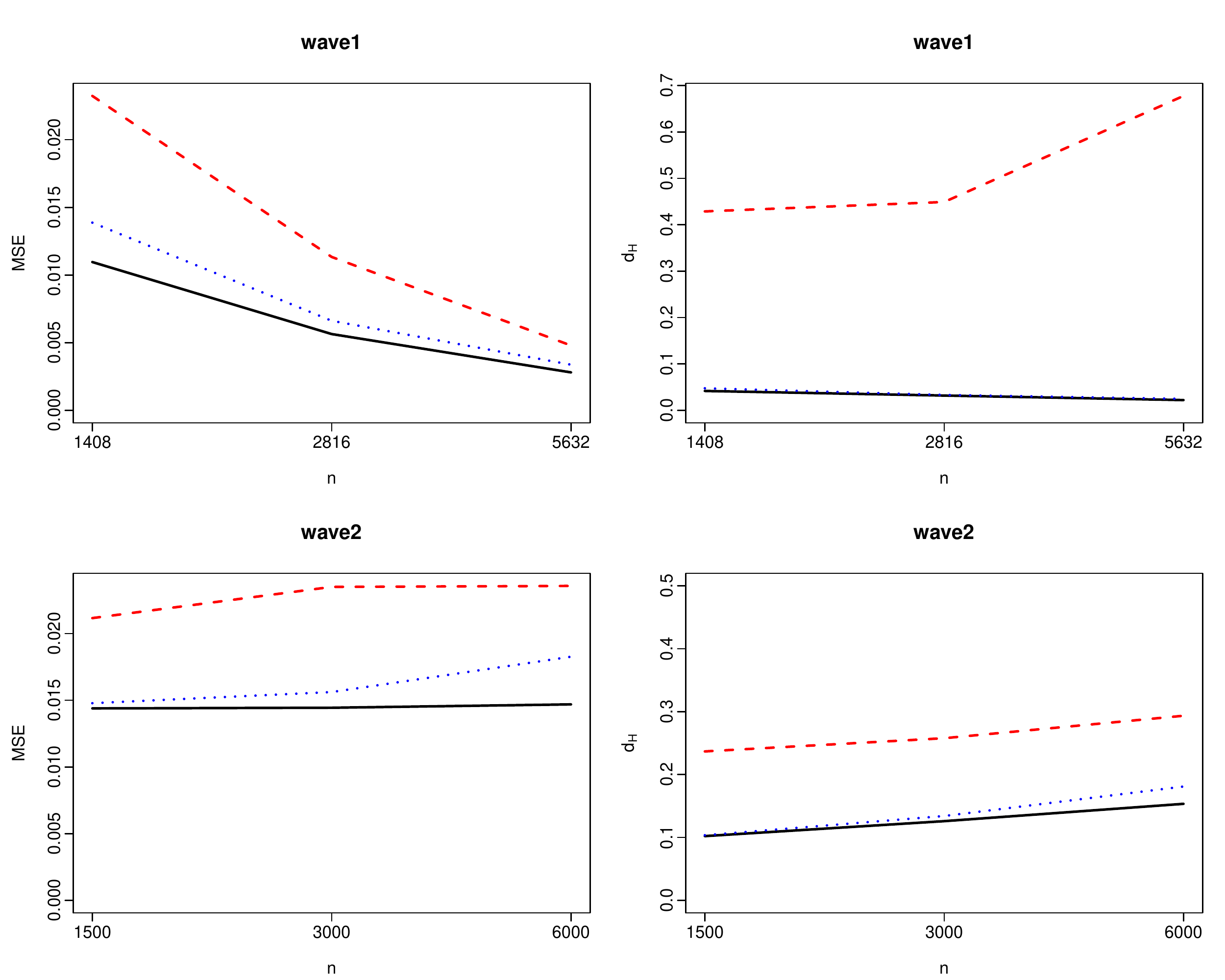}
 \end{center}
 \caption{\label{Fig:wave_results}
 Results for CPOP (black solid line), NOT (blue dotted line) and trend-filtering (red dashed line) for {\texttt{wave1}} (top row) and {\texttt{wave2}} (bottom row). We give results for mean square error of the 
 estimate of the mean (left-hand column) and for the accuracy of the estimates of the changepoint locations, measured via $d_H$ (right-hand column). For {\texttt{wave1}} we considered data sets of length $n=1408$,
  $n=2816$ and $n=5632$, each data set having 8 segments. For {\texttt{wave2}} we considered data sets of length $n=1500$, $n=3000$ and $n=6000$, each data set having segments of length 150. Results are averaged over 
  100 data sets in for each scenario and each value of $n$.
 }
\end{figure}

First we analyse data from the {\texttt{wave1}} and {\texttt{wave2}} scenarios. We consider different lengths of data with either a fixed number of changepoints ({\texttt{wave1}}) or with the number of changepoints increasing linearly with
the number of data points ({\texttt{wave2}}). For both {\texttt{wave1}} and {\texttt{wave2}} there is a substantial change in the slope of the mean at each changepoint. As such, these represent relatively straightforward
scenarios for detecting changepoints, and both NOT and CPOP perform well at detecting the number of changepoints: NOT correctly identifies the number of changepoints for all 600 simulated data sets, and CPOP correctly
identifies the number of changepoints in over 99\% of these cases. By comparison trend-filtering substantially over-estimates the number of changepoints in all cases. For {\texttt{wave1}} the average number of changes
detected is 16 for $n=1408$, rising to 29 for $n=5632$, when the true number of changes is 7. We have similar over-estimation for {\texttt{wave2}}. The reason for this is the use the $L_1$ penalty for the change in slope. The $L_1$ penalty is the same for multiple
consecutive changes in slope of the same sign as it is for one large change in slope. As a result trend-filtering tends to introduce multiple changepoints around each actual change.  

This over-estimation of the number of changes results in the much larger value of $d_H$ for this method than for NOT and CPOP: see the right-hand plots of Figure \ref{Fig:wave_results}.
Whilst NOT and CPOP perform similarly in terms of accuracy when estimating changepoint location, CPOP is more accurate in terms of estimating the underlying mean: see the MSE results in the left-hand plots of
Figure \ref{Fig:wave_results}. Again both methods perform better than trend-filtering. We believe the reason for this is that trend-filtering shrinks the change in slope towards 0. For signals like {\texttt{wave1}}
and {\texttt{wave2}} where all changes in slope are substantial, this causes trend-filtering to under-estimate these changes. This can introduce substantial error at estimating the mean in regions around each
changepoint.

\begin{figure}
 \begin{center}
  \includegraphics[scale=0.6]{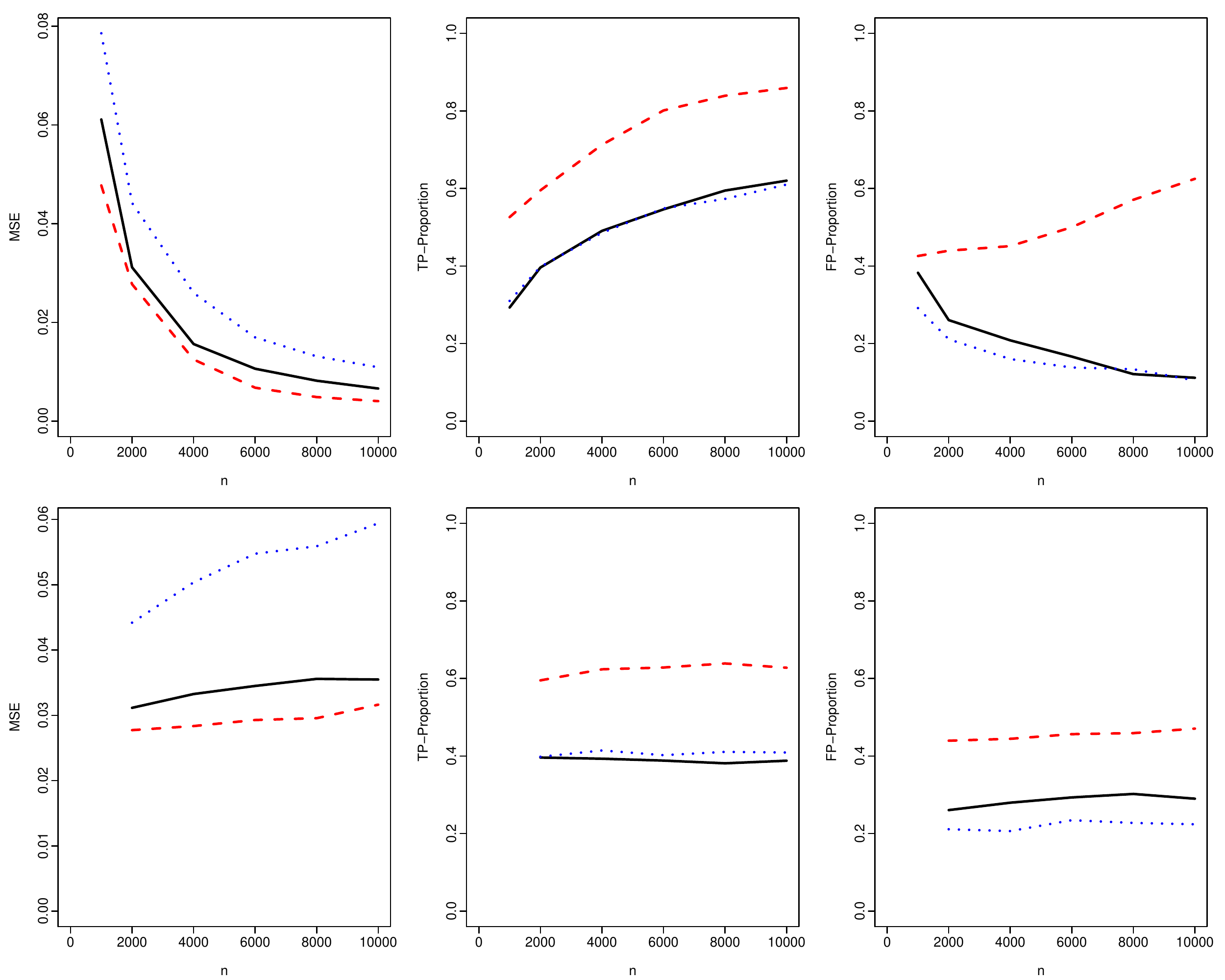}
 \end{center}
 \caption{\label{Fig:Random_results}
 Results for CPOP (black solid line), NOT (blue dotted line) and trend-filtering (red dashed line) for the {\texttt{Random}} scenario with a fixed number of  changepoints (top row) and a fixed segment length (bottom row). We give results for mean square error of the 
 estimate of the mean (left-hand column) and for the accuracy of the estimates of the changepoint locations, measured via the proportion of true-positives (middle column) and of false-positives (right-hand column). For the top row 
 we have 20 segments for each data set, for the bottom row we have segments of length 100 for each data set.  Results are averaged over 100 data sets for each case and each value of $n$.
 }
\end{figure}

We now compare the three methods on the {\texttt{Random}} simulation scenario. We consider data sets of length varying from 1000 to 10000, with either a fixed number of 20 segments or with the segment length
fixed to 100. This is a harder scenario than either {\texttt{wave1}} or {\texttt{wave2}} as the change in slope differs considerably from changepoint to changepoint, with the change in slope being small in many cases
(see the example data sets in the bottom row of Figure \ref{Fig:Scenarios}). As a result there are many changepoints that are hard to detect. In all cases CPOP and NOT under estimate the number of changes, while 
trend-filtering still over estimates this number. These two different sources of error are masked in the measure $d_H$, and thus we summarise the accuracy of changepoint detection through true-positive and false-positive
proportions. To calculate these we say that an actual change is detected if there is an estimated changepoint within a certain distance of it. The results we show have set this distance to be a fifth of the segment length,
though qualitatively similar results are obtained with different choices. We calculate the number of false positives as the number of changepoints detected less the number of true positives. Our results
are in terms of the true-positive proportion, which is the proportion of actual changepoints detected, and the false-positve proportion, the proportion of detected the changepoints that are false-positive.

Results are shown in Figure \ref{Fig:Random_results}. These are qualitatively different from the earlier results. For this problem we see that trend-filtering is most accurate in terms of estimating the underlying mean.
We believe that trend-filtering is more suited to this scenario as there are a range of values for how much the slope changes at each changepoint, including many cases where the change is small. Hence the
shrinking of the change in slope that trend-filtering induces is actually beneficial. As trend-filtering estimates more changes, it detects a higher proportion of true changepoints, but it has a high false-positive proportion:
in all cases over 40\% of the changepoints it finds are false-positives. By comparison both NOT and CPOP have lower false positive proportions, and encouragingly, this proportion decreases as the segment length
increases (see top right-hand plot in Figure \ref{Fig:Random_results}). Whilst NOT is marginally better in terms of accuracy of the detected changepoints, CPOP is substantially more accurate in terms of its estimate
of the underlying mean.

\section{Bacterial Flagella Motor Data} \label{sec:bacteria}

\begin{figure}
 \begin{center}
  \includegraphics[scale=0.7,angle=0]{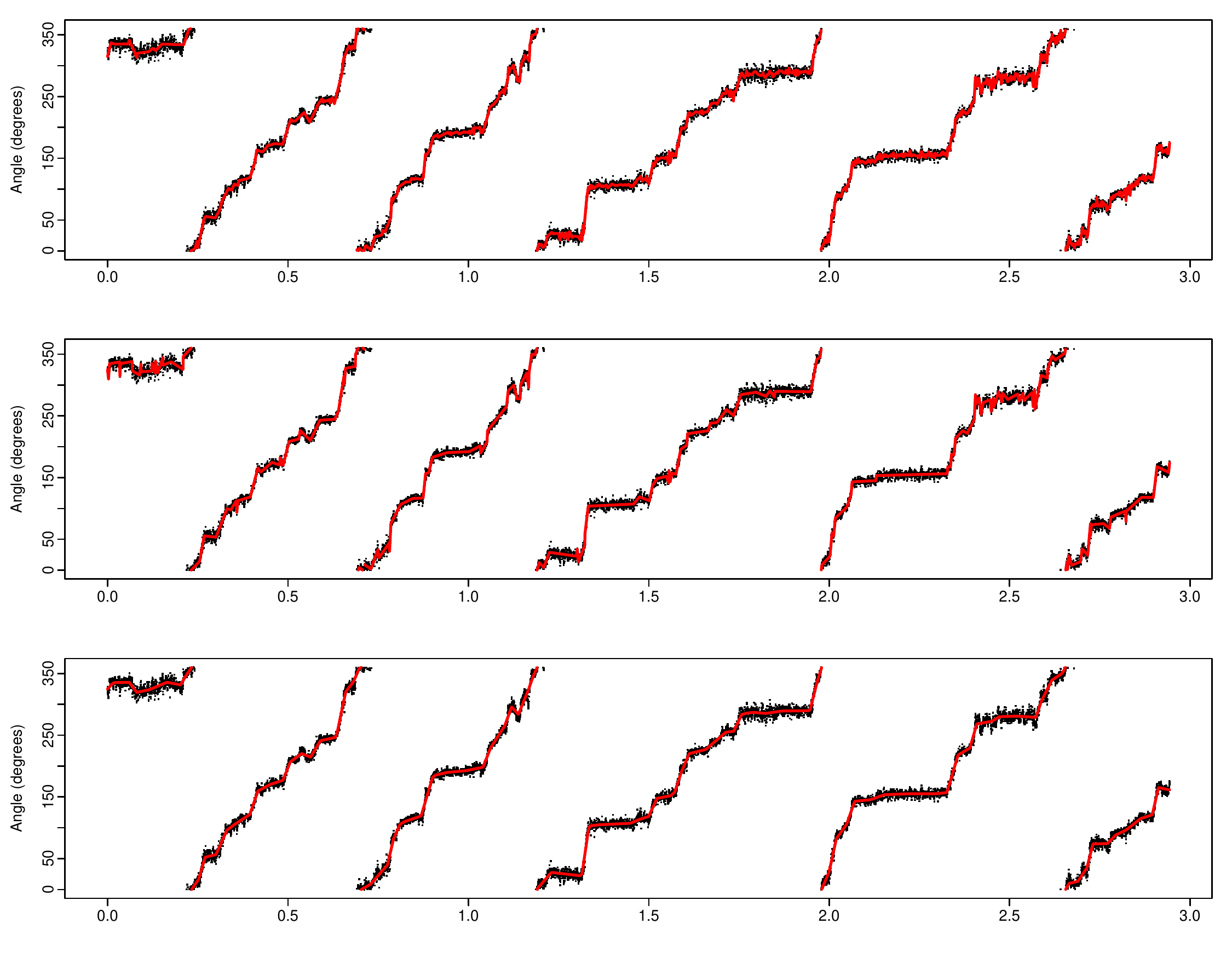}
 \end{center}
\caption{\label{Fig:Sowa2} Time-series of angular position \cite[data from][]{sowa} and example fits obtained by NOT (top); CPOP (middle) and trend-filtering (bottom). The fit obtained by CPOP is the one
that contains the same number, 182, of changepoints as that found by NOT (see text for more details). For ease of presentation we have plotted the angle of the bacteria, the model we fit assumes continuity of angles of
360 degrees (top of each plot) and 0 degrees (bottom of each plot).}
\end{figure}

We return to the bacterial flagella motor data we introduced in Section \ref{sec:introduction} and Figure \ref{Fig:Sowa1}. For more background on these biological systems see \citep{sowa,Sowa2008,Zhou1998}. 
Data similar to those we analyse has been collected by \cite{ryu2000torque,chen2000solvent,chen2000torque,sowa2003torque} among others. Here we look at how well we can extract the angular motion
by fitting changepoint models, and in particular change-in-slope models using the CPOP algorithm. The data we analyse comes from \cite{sowa} and is shown in Figure \ref{Fig:Sowa2}. It consists of 11,912 observations.

\begin{figure}
 \begin{center}
  \includegraphics[scale=0.4,angle=0]{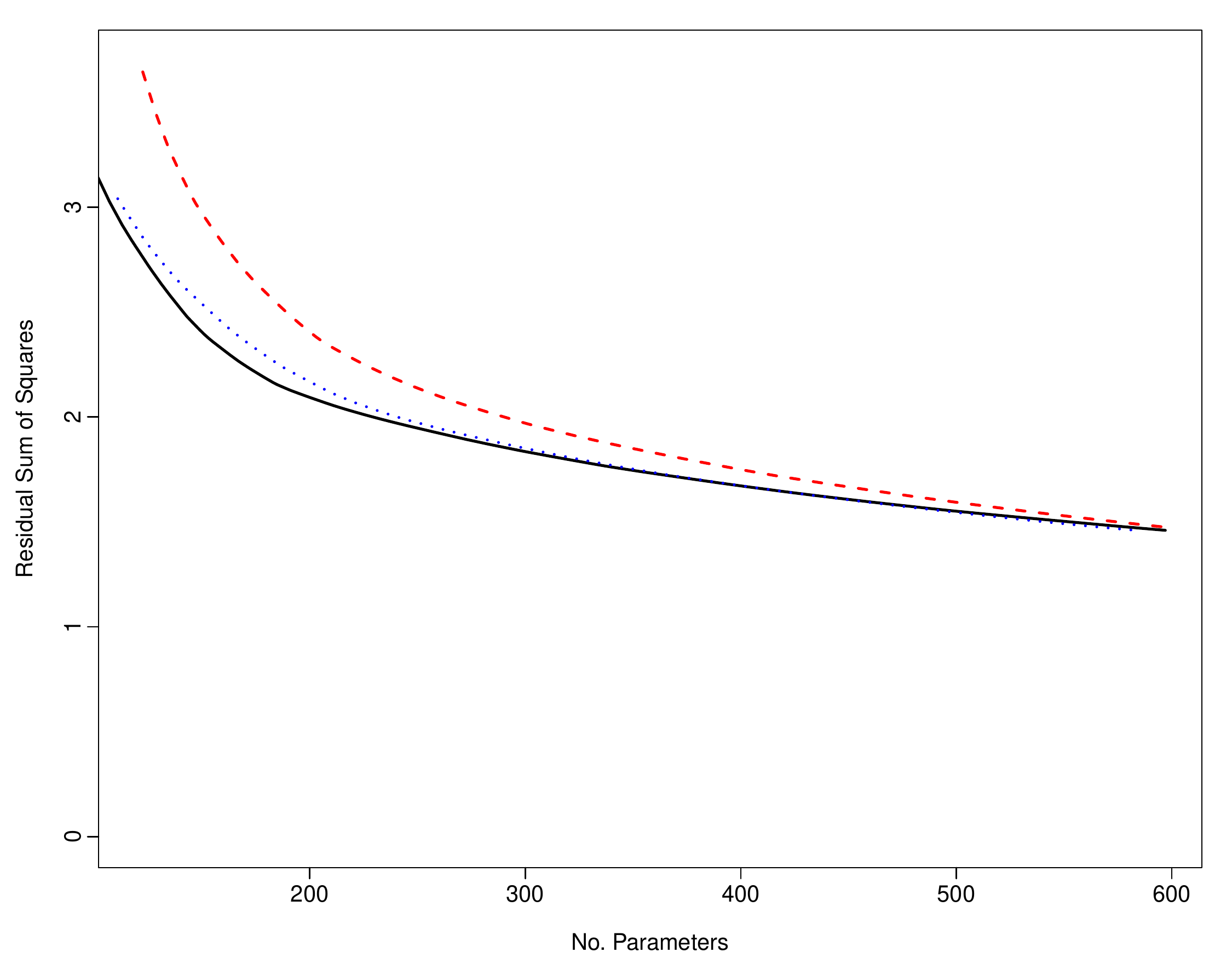}
 \end{center}
\caption{\label{Fig:Sowa_comp} Accuracy of fits of data shown in Figure \ref{Fig:Sowa2} by a piecewise constant mean (red dashed line), a continuous piecewise-linear mean (black full line) and a discontinuous piecewise-linear 
mean (blue dotted line). For each type of line we
found the best segmentation, in terms of minimising the residual sum of squares (RSS) of the fit, for a range of the number of changepoints. We plot the RSS against the number of free parameters of the fitted mean
function for each case.
}
\end{figure}

The aim of our analysis is to fit the underlying angular position. We first compared fitting a continuous piecewise linear
mean to both fitting a piecewise constant mean and a discontinuous piecewise linear mean. We fit the latter two by minimising the residual sum of squares plus a penalty times the number of changepoints, 
using the PELT algorithm \cite[]{Killick2012a}. In all cases
we varied the penalty value using the CROPS algorithm \cite[]{Haynes2014}. Different penalty values lead to optimal segmentations with different numbers of changepoints. For each different segmentation we calculated the 
actual residual sum of squares of the fit we obtained. A plot of this against the number of free parameters in the fitted mean is shown in Figure \ref{Fig:Sowa_comp}. We can see that fitting a continuous piecewise-linear
function, which is more natural for this application, leads to a uniformly better fit to the data than the change in mean for any given number of parameters. The assumption of continuity also  gives improvements for
fitted means with fewer than 400 parameters. While the differences in residual sum of squares looks small, due to the large number of observations, the reduction in log-likelihood, under a model where the residuals are 
iid Gaussian, is still substantial. For example, for models with fewer than 350 parameters, the best fitting continuous mean has a log-likelihood that is 32.4 units greater than the best fitting discontinuous mean.

We also compared the accuracy of using CPOP to analyse this data to that of using NOT and trend-filtering. A comparison of the fits obtained using NOT, CPOP and trend-filtering are shown in Figure \ref{Fig:Sowa2}. 
We ran  NOT with a total of $10^6$ random intervals, and have plotted the segmentation that minimised the SIC. This segmentation has 794 changepoints, largely because it substantially overfits the latter part of the
data. For comparison, an example fit from CPOP is also shown. The segmentation obtained using CPOP has 182 changepoints. Despite fewer changes, it has a smaller residual sum of squares than the segmentation that NOT found:
1.72 as compared to 1.80.


We also ran trend-filtering for a range of penalty values. For all penalty values that gave a reasonable fit to the data, the number of changes in slope was large: with changes at more than half the time-points. 
In these cases the majority of changes in slope with small. As a crude approach to choosing a sensible segmentation we defined there to be a change-point if the change in slope was non-zero after rounding to 3 
decimal places. Using this definition we then found the segmentation that minimised the SIC. This is shown in the bottom plot of \ref{Fig:Sowa_comp}. This had 278 changepoints under our definition, 
and 10,850 actual changes in slope. We see that the estimated mean we obtained appears to under-fit the data in a number of places. It has a higher residual sum of squares, 2.94, than the fitted mean shown for either  
CPOP or NOT.

\section{Discussion} \label{sec:discussion}

We have presented a continuous-state dynamic programming algorithm for finding the best continuous piecewise linear fit to data under a criterion that measures fit to the data using the residual sum of squares and penalises complexity
through the number of changes in slope. This is a setting where standard dynamic programming approaches for changepoint detection do not work, due to the dependence across segments imposed by the continuity constraint.
Empirically this approach is feasible for data with up to 10,000 data points and 100s of changepoints. For such challenging scenarios, we see from the analysis of the bacterial flagella motor data, that this
method can produce a substantially better fit to the data than faster approximate alternatives like NOT and trend-filtering. 

The dynamic programming approach we have used has the potential to be applied to a much wider range of changepoint problems with dependence across segments. The key requirement is that we can construct a recursion for
a set of functions, our $f^{t}(\phi)$, that are piecewise quadratic in some univariate parameter $\phi$. This requires that we measure fit to the data through the residual sum of squares, that the dependence of the parameters
in successive segments is through a univariate quantity $\phi$, and that any constraints on parameters in successive segments respect the piecewise quadratic nature of $f^{t}(\phi)$. This would cover change in mean 
or slope under monotonicity constraints, our change in slope model with an additional $L_1$ or $L_2$ penalty on the change in slope, or more general models for the mean that are piecewise polynomonial and continuous.

The requirement that dependence across segments is through a univariate quantity comes from our functional pruning approach. Such pruning is important for reducing the computational complexity of the algorithm. It is
unclear whether functional pruning can be implemented for piecewise quadratic functions, $f^{t}(\phi)$, when $\phi$ is not univariate as the line search approach we take does not generalise beyond the univariate case. 
Even if not, it may be possible to develop efficient algorithms that implement an approximate version of functional pruning.

{\bf Acknowledgements} 
This work was supported by EPSRC grants EP/N031938/1 (StatScale) and EP/H023151/1 (STOR-i). We thank Ashley Nord and Richard Berry for helpful discussions on the analysis of the bacterial flagella motor data; and
Rafal Baranowksi, Yining Chen and Piotr Fryzlewicz for advice on using NOT.

\bibliography{refs}
\bibliographystyle{royal}


\begin{appendices}

\section{Updates for Quadratic Functions}

\label{sec:coupdate}

In Section \ref{sec:methodname} (equation \ref{eq:1}) we define a function, $f_{\boldsymbol{\tau}}^t(\phi)$, as the minimum cost of segmenting $\mathbf{y}_{1:t}$ with changepoints at $\boldsymbol{\tau}=\tau_1,\hdots,\tau_k$ and fitted value $\phi_t=\phi$ at time $t$. We then derived a recursion for these functions as follows
\begin{align}
  \label{eq:A23} f_{\boldsymbol{\tau}}^t(\phi)=\min_{\phi'}\left\{f_{\tau_1,\hdots,\tau_{k-1}}^{\tau_k}(\phi')+\mathcal{C}(y_{\tau_k+1:t},\phi',\phi)+\beta+h(\tau_{i+1}-\tau_i)\right\}.
\end{align}


The functions $f_{\boldsymbol{\tau}}^t(\phi)$ are quadratics in $\phi$, and we denote $f_{\boldsymbol{\tau}}^t(\phi)$ as follows
\begin{align}
  f_{\boldsymbol{\tau}}^t(\phi)=a_{\boldsymbol{\tau}}^t+b_{\boldsymbol{\tau}}^t\phi+c_{\boldsymbol{\tau}}^t\phi^2, \label{quadcost}
\end{align}
for some constants $a_{\boldsymbol{\tau}}^t$, $b_{\boldsymbol{\tau}}^t$ and $c_{\boldsymbol{\tau}}^t$. We then wish to calculate these coeffcients by updating the coefficients 
that make up $f_{\tau_1,\hdots,\tau_{k-1}}^{\tau_k}(\phi')$ using (\ref{eq:A23}). To do this we need to write the cost for the segment from $\tau_k+1$ to $t$ in quadratic form. 
Defining the length of the segment as $s=t-\tau_k$ this cost can be written as
\begin{align} \mathcal{C}(\mathbf{y}_{\tau_k+1:t},\phi',\phi)=&\frac{(s+1)(2s+1)}{6s\sigma^2}\phi^2+\left(\frac{(s+1)}{\sigma^2}-\frac{(s+1)(2s+1)}{3s\sigma^2}\right)\phi'\phi\notag\\
&-\left(\frac{2}{s\sigma^2}\sum y_j(j-\tau_k)\right)\phi+\left(  
\frac{1}{\sigma^2}\sum y_i^2\right)\notag\\
&+2\left(\frac{1}{s\sigma^2}\sum y_j(j-\tau_k)-\frac{1}{\sigma^2}\sum y_i\right)\phi'+\frac{(s-1)(2s-1)}{6s\sigma^2}\phi'^2.\label{segquad}
\end{align}

Writing (\ref{segquad}) as $A\phi^2+B\phi'\phi+C\phi+D+E\phi'+F\phi'^2$ for constants $A$, $B$, $C$, $D$ and $E$, substituting (\ref{segquad})  into (\ref{eq:A23}) 
and minimising out $\phi'$ we can get the formula for the updating the coefficients of the quadratic $f_{\boldsymbol{\tau}}^t(\phi)$:

  \begin{align}
    a_{\boldsymbol{\tau}}^t&=A-\frac{B^2}{4\left(a^{\tau_k}_{(\tau_1,\hdots,\tau_{k-1})}+F\right)},\notag\\
    b_{\boldsymbol{\tau}}^t&=C-\frac{\left(b^{\tau_k}_{(\tau_1,\hdots,\tau_{k-1})}+E\right)B}{2\left(a^{\tau_k}_{(\tau_1,\hdots,\tau_{k-1})}+F\right)},\notag\\
    c_{\boldsymbol{\tau}}^t&=c^{\tau_k}_{(\tau_1,\hdots,\tau_{k-1})} + D- 
    \frac{\left(b^{\tau_k}_{(\tau_1,\hdots,\tau_{k-1})}+E\right)^2}{4\left(a^{\tau_k}_{(\tau_1,\hdots,\tau_{k-1})}+F\right)}+\beta+h(t-\tau_k).\label{eq:Aupdates}
  \end{align}

\section{Proofs}
\label{App:Proof}

\subsection{Proof of Theorem~\ref{thr:fp}}

The proof of Theorem~\ref{thr:fp} works by contrapositive. We show that if $(\boldsymbol{\tau},s)\in\overset{*}{\mathcal{T}}_t$ then a necessary condition of this is that $\boldsymbol{\tau}\in\overset{*}{\mathcal{T}}_s$, taking the contrapositive of this gives Theorem~\ref{thr:fp}.

\begin{proof}
  Assume $(\boldsymbol{\tau},s)\in\overset{*}{\mathcal{T}}_t$, then there exists $\phi$ such that
\begin{align}
  f^t(\phi)&=f_{(\boldsymbol{\tau},s)}^t(\phi),\notag
\end{align}
Now for any $\phi^*$,
\begin{align}
f^s(\phi^*)+\mathcal{C}(\mathbf{y}_{s+1:t},\phi^*,\phi)+\beta&\geq\min_{\phi',r}\left[f^r(\phi')+\mathcal{C}(\mathbf{y}_{r+1:t},\phi',\phi)+\beta\right],\notag\\
&=f^t(\phi),\notag\\
&=f_{(\boldsymbol{\tau},s)}^t(\phi),\notag\\
&=\min_{\phi''}\left\{f_{\boldsymbol{\tau}}^s(\phi'')+\mathcal{C}(\mathbf{y}_{s+1:t},\phi'',\phi)+\beta\right\},\label{thisone}\\
&=f_{\boldsymbol{\tau}}^s(\phi^A)+\mathcal{C}(\mathbf{y}_{s+1:t},\phi^A,\phi)+\beta,\notag
\end{align}
where $\phi^A$ is the value of $\phi''$ which minimises (\ref{thisone}). As $\phi^*$ can be chosen as any value, we can choose it as $\phi^A$. By cancelling terms we get $f^s(\phi^A)\geq f_{\boldsymbol{\tau}}^s(\phi^A)$ and hence (from (\ref{eq:4})), $f^s(\phi^A)= f_{\boldsymbol{\tau}}^s(\phi^A)$ and therefore $\boldsymbol{\tau}\in\overset{*}{\mathcal{T}}_s$.
We have shown that if $(\boldsymbol{\tau},s)\in\overset{*}{\mathcal{T}}_t$ then $\boldsymbol{\tau}\in\overset{*}{\mathcal{T}}_s$, by taking the contrapositive the theorem holds.
\end{proof}

\subsection{Proof of Theorem~\ref{thr:3}}

The proof for Theorem~\ref{thr:3} follow a similar argument to the corresponding proof in \cite{Killick2012a}. 
However we have to add a segment consisting of the single point $y_{t+1}$ to deal with the dependence between the segments.

\begin{proof}
  \label{sec:condition-that-costmbic}
  Let $\boldsymbol{\tau}^*$ denote the optimal segmentation of $\mathbf{y}_{1:t}$.
First consider $T=t+1$. As adding a changepoint without penalty will always reduce the cost, it is straightforward to show
\begin{align}
   f^T_{\boldsymbol{\tau}}(\phi)&\geq \min_{\phi'}\left[f^t_{\boldsymbol{\tau}}(\phi')+\mathcal{C}(y_{t+1},\phi',\phi)\right],\notag\\
   &=\min_{\phi'}[f_{\boldsymbol{\tau}}^t(\phi')] + \min_{\phi'}[\mathcal{C}(y_{t+1},\phi',\phi)],\notag\\
&> \min_{\phi'}\left[f^{t}(\phi')\right]+K+\min_{\phi'}[\mathcal{C}(y_{t+1},\phi',\phi)],\notag\\
 &\geq  \min_{\phi'}\left[f^{t}(\phi') + \mathcal{C}(y_{t+1},\phi') + \beta+h(1) \right].\notag
\end{align}
Thus segmenting $\mathbf{y}_{1:T}$ with changepoints $\boldsymbol{\tau}$ always has a greater cost than segmenting $\mathbf{y}_{1:T}$
with changepoints $(\boldsymbol{\tau}^*,t)$.

Now we consider $T>t+1$. We start by noting 
that by adding changes, at any point, without the penalty term and minimising over the corresponding $\phi$ values will also decrease the cost. Therefore we have
\begin{align}
  \label{eq:5}
  f^T_{\boldsymbol{\tau}}(\phi)\geq \min_{\phi',\phi''}\left[f^t_{\boldsymbol{\tau}}(\phi')+\mathcal{C}(y_{t+1},\phi',\phi'')+\mathcal{C}(\mathbf{y}_{t+2:T},\phi'',\phi)\right].
\end{align}

Then assuming that \eqref{eq:hold} is true, it can be shown that the segmenting the data $\mathbf{y}_{1:T}$ with changepoints $\boldsymbol{\tau}$ is always sub-optimal.

So from (\ref{eq:5}) and using (\ref{eq:hold}),
\begin{align}
   f^T_{\boldsymbol{\tau}}(\phi)&\geq \min_{\phi',\phi''}\left[f^t_{\boldsymbol{\tau}}(\phi')+\mathcal{C}(y_{t+1},\phi',\phi'')+\mathcal{C}(\mathbf{y}_{t+2:T},\phi'',\phi)\right],\notag\\
   &\geq\min_{\phi'}[f_{\boldsymbol{\tau}}^t(\phi')] + \min_{\phi',\phi''}[\mathcal{C}(y_{t+1},\phi',\phi'') + \mathcal{C}(\mathbf{y}_{t+2:T},\phi'',\phi)],\notag\\
&> \min_{\phi'}\left[f^{t}(\phi')\right]+K+\min_{\phi',\phi''}[\mathcal{C}(y_{t+1},\phi',\phi'') + \mathcal{C}(\mathbf{y}_{t+2:T},\phi'',\phi)],\notag\\
 &\geq  \min_{\phi',\phi''}\left[f^{t}(\phi') + \mathcal{C}(y_{t+1},\phi',\phi'') + \beta+h(1) + \mathcal{C}(\mathbf{y}_{t+2:T},\phi'',\phi)+\beta+h(T-t+1)\right].\notag
\end{align}
The last step  is due to the cost on a single point, $\mathcal{C}(y_t,\phi',\phi'')$ only depending on $\phi''$, and by using the definition of $K$.

Therefore the cost of segmenting $\mathbf{y}_{1:T}$ with changepoints $\boldsymbol{\tau}$ is always greater than the cost of segmenting $\mathbf{y}_{1:T}$
with changepoints $(\boldsymbol{\tau}^*,t,t+1)$ (where $\boldsymbol{\tau}^*$ is the optimal segmentation of $\mathbf{y}_{1:t}$) and this holds for all $T>t+1$ and hence $\boldsymbol{\tau}$ can be pruned.
\end{proof}

\section{Pseudo-Code for CPOP}
\label{App:Alg}

\IncMargin{1em}
\begin{algorithm}[!h]
\SetKwData{Left}{left}\SetKwData{This}{this}\SetKwData{Up}{up}
\SetKwFunction{Union}{Union}\SetKwFunction{FindCompress}{FindCompress}
\SetKwInOut{Input}{Input}\SetKwInOut{Output}{Output}
\Input{Set of data of the form $\mathbf{y}_{1:n}=(y_1,\hdots,y_n)$.\\
A positive penalty constant, $\beta$, and a non-negative, non-decreasing penalty function $h(\cdot)$.}
\BlankLine
Let $n=$ length of data\;
set $\hat{\mathcal{T}}_1=\{0\}$\;
and set $K=2\beta+h(1)+h(n)$\;
\For{$t=1,\hdots,n$}{
  \For{$\boldsymbol{\tau}\in\hat{\mathcal{T}}_t$}{
   \If{$\tau=\{0\}$}{
     $f_{\boldsymbol{\tau}}^t(\phi)=\displaystyle\min_{\phi'}\mathcal{C}(\mathbf{y}_{1:t},\phi',\phi)+h(t)$\;
    }
    \Else{
     $f_{\boldsymbol{\tau}}^t(\phi)=\displaystyle\min_{\phi'}\left\{f_{\tau_1,\hdots,\tau_{k-1}}^{\tau_k}(\phi')+\mathcal{C}(\mathbf{y}_{\tau_k+1:t},\phi',\phi)+h(t-\tau_k)+\beta\right\}$\;
    }
 }
\For{$\boldsymbol{\tau}\in\hat{\mathcal{T}}_t$}{
      $Int^t_{\boldsymbol{\tau}}=\left\{\phi:f_{\boldsymbol{\tau}}^t(\phi)=\displaystyle\min_{\boldsymbol{\tau}'\in\hat{\mathcal{T}}_t}f_{\boldsymbol{\tau}'}^t(\phi)\right\}$\;
$\overset{*}{\mathcal{T}}_t=\left\{\boldsymbol{\tau}: Int^t_{\boldsymbol{\tau}}\neq\emptyset\right\}$\;
      $\hat{\mathcal{T}}_{t+1}=\hat{\mathcal{T}}_t\cup\left\{(\boldsymbol{\tau},t):\boldsymbol{\tau}\in\overset{*}{\mathcal{T}}_t\right\}$\;
    }
$\hat{\mathcal{T}}_{t+1}=\left\{\boldsymbol{\tau}\in\hat{\mathcal{T}}_{t+1}:\displaystyle\min_{\phi} f_{\boldsymbol{\tau}}^t(\phi)\leq\displaystyle\min_{\phi',\boldsymbol{\tau}'}
\left[f_{\boldsymbol{\tau}'}^t(\phi')\right]+K\right\}$\;
}
$f_{opt}=\displaystyle\min_{\boldsymbol{\tau},\phi}f^n_{\boldsymbol{\tau}}(\phi)$\;
$\boldsymbol{\tau}_{opt}=\displaystyle\operatorname*{arg\,min}_{\boldsymbol{\tau}}\left[\displaystyle\min_{\phi}f^n_{\boldsymbol{\tau}}(\phi)\right]$\;
\Output{The optimal cost, $f_{opt}$, and the corresponding changepoint vector, $\boldsymbol{\tau}_{{opt}}$.}
\BlankLine
\caption{Algorithm for Continuous Piecewise-linear Optimal Partitioning (CPOP)}\label{algo_optpiecewise}
\end{algorithm}\DecMargin{1em}

The CPOP algorithm uses Algorithm ~\ref{algo_calcsettau} to calculate the intervals on which each function is optimal. This then enables the functions that are not optimal for any value of $\phi$ to be removed. The idea of this algorithm is as follows.

We initialise the algorithm by setting the current parameter value as $\phi_{curr}=-\infty$ and comparing the cost functions in our current set of candidates (which we initialise as $\mathcal{T}_{temp}=\hat{\mathcal{T}}_t$) to get the optimal segmentation for this value, $\boldsymbol{\tau}_{curr}$. For each $\boldsymbol{\tau}\in\mathcal{T}_{curr}$  we calculate where $f^t_{\boldsymbol{\tau}}$ next intercepts with $f^t_{\boldsymbol{\tau}_{curr}}$ (smallest value of $\phi$ for which $f^t_{\boldsymbol{\tau}}(\phi)=f^t_{\boldsymbol{\tau}_{curr}}(\phi)$ and $\phi>\phi_{curr}$) and store this as $x_{\boldsymbol{\tau}}$. If for a $\boldsymbol{\tau}\in\mathcal{T}_{temp}$ we have $x_{\boldsymbol{\tau}}=\emptyset$ (i.e.  $f^t_{\boldsymbol{\tau}}$ doesn't intercept with $f^t_{\boldsymbol{\tau}_{curr}}$ for any $\phi>\phi_{curr}$) then we remove $\boldsymbol{\tau}$ from $\mathcal{T}_{temp}$. We take the minimum of $x_{\boldsymbol{\tau}}$ (the first of the intercepts)  and set it as our new $\phi_{curr}$ and the corresponding changepoint vector that produces it as $\boldsymbol{\tau}_{curr}$. We repeat this procedure until the set $\mathcal{T}_{temp}$ consists of only a single value $\boldsymbol{\tau}_{curr}$ which is the optimal segmentation for all future $\phi>\phi_{curr}$.

\IncMargin{1em}
\begin{algorithm}
\SetKwData{Left}{left}\SetKwData{This}{this}\SetKwData{Up}{up}
\SetKwFunction{Union}{Union}\SetKwFunction{FindCompress}{FindCompress}
\SetKwInOut{Input}{Input}\SetKwInOut{Output}{Output}
\Input{Set of changepoint candidate vectors $\hat{\mathcal{T}}_t$ for current timestep, $t$,\\
Optimal segmentation functions $f_{\boldsymbol{\tau}}^t(\phi)$ for current time step $t$ and $\boldsymbol{\tau}\in\hat{\mathcal{T}}_t$.}
\BlankLine
$\mathcal{T}_{temp}=\hat{\mathcal{T}}_t$\;
$Int^t_{\boldsymbol{\tau}}=\emptyset$ for $\boldsymbol{\tau}\in\hat{\mathcal{T}}_t$\;
$\phi_{curr}=-\infty$\;
$\boldsymbol{\tau}_{curr}=\displaystyle\operatorname*{arg\,min}_{\boldsymbol{\tau}\in\mathcal{T}_{temp}}\left[f_{\boldsymbol{\tau}}^t(\phi_{curr})\right]$\;
\While{$\mathcal{T}_{temp}\backslash\{\boldsymbol{\tau}_{curr}\}\neq\emptyset$}{
\For{$\boldsymbol{\tau}\in\mathcal{T}_{temp}\backslash\{\boldsymbol{\tau}_{curr}\}$}{
$x_{\boldsymbol{\tau}}=\min\{\phi:f_{\boldsymbol{\tau}}^t(\phi)-f_{\boldsymbol{\tau}_{curr}}^t(\phi)=0 \And \phi>\phi_{curr}\}$\;
\If{$x_{\boldsymbol{\tau}}=\emptyset$}{
$\mathcal{T}_{temp}=\mathcal{T}_{temp}\backslash\{\boldsymbol{\tau}\}$
}

}
$\boldsymbol{\tau}_{new}=\displaystyle\operatorname*{arg\,min}_{\boldsymbol{\tau}}(x_{\boldsymbol{\tau}})$\;
$\phi_{new}=\displaystyle\min_{\boldsymbol{\tau}}(x_{\boldsymbol{\tau}})$\;
$Int^t_{\boldsymbol{\tau}_{curr}}=[\phi_{curr},\phi_{new}]\cup Int^t_{\boldsymbol{\tau}_{curr}}$\;
$\boldsymbol{\tau}_{curr}=\boldsymbol{\tau}_{new}$\;
$\phi_{curr}=\phi_{new}$\;
}
\Output{The intervals $Int^t_{\boldsymbol{\tau}}$ for $\boldsymbol{\tau}\in\hat{\mathcal{T}}_t$}
\BlankLine
\caption{Algorithm for calculation of $Int^t_{\boldsymbol{\tau}}$ at time $t$}\label{algo_calcsettau}
\end{algorithm}\DecMargin{1em}

\end{appendices}
\end{document}

%% file: header.tex
\usepackage{graphicx}
\usepackage{natbib}
\usepackage{amsmath}
\usepackage{amsfonts}
\usepackage{url}

\newenvironment{proof}{\begin{trivlist} \item[]
{\bf Proof.}}{\nolinebreak
\hfill \rule{2mm}{2mm} \end{trivlist}}

\newcounter{ctr}

\newcounter{ctr1}

\newcounter{ctr2}

\newcounter{ctr3}

\newtheorem{definition}{Definition}[section]    
\newtheorem{theorem}[definition]{Theorem}
\newenvironment{theorem*}[1]{{\bf Theorem #1} \begin{itshape}}{\end{itshape}}

\newenvironment{corollary*}[1]{{\bf Corollary #1} \begin{itshape}}{\end{itshape}}

\newenvironment{proposition*}[1]{{\bf Proposition #1} \begin{itshape}}{\end{itshape}}

\newcommand{\ud}{\, {\rm d} \kern-.015em }

\newcommand{\modulus}[1]{\left| \kern.05em #1 \kern.05em \right|}
\newcommand{\norm}[1]{\left\| \kern.05em #1 \kern.05em \right\|}
\newcommand{\inner}[1]{\left\langle \kern.05em #1 \kern.05em \right\rangle }

\newcommand{\pick}[2]{\renewcommand{\arraystretch}{0.6}
\left( \kern-.4em \begin{array}{c} #1 \\ #2 \end{array} \kern-.4em \right) }

